\documentclass[twocolumn,preprintnumbers,amsmath,amssymb]{revtex4}
\usepackage{graphicx}
\usepackage{amsmath}
\usepackage[dvipsnames]{color}
\usepackage{bm}

\usepackage[colorlinks=true,citecolor=blue]{hyperref}
\hypersetup{colorlinks=true,citecolor=blue,linkcolor=red,urlcolor=blue}

\usepackage{color}

\def \brho{{\hbox{\boldmath $\rho$}}}

\begin{document}

\title{{\em Ab initio} study of electromagnatic modes in two-dimensional semiconductors: Application to doped phosphorene}

\author{Dino Novko$^{1,2}$, Keenan Lyon$^{3}$, Duncan J. Mowbray$^{4}$, Vito Despoja$^{1,2}$}
\affiliation{$^1$Institute of Physics, 10000 Zagreb, Croatia}
\affiliation{$^2$Donostia International Physics Center (DIPC), 20018 San Sebastián, Spain}
\affiliation{$^3$Department of Applied Mathematics, University of Waterloo, Waterloo, Ontario, Canada}
\affiliation{$^3$Department of Physics and Astronomy, Uppsala University, S-751 20, Uppsala, Sweden}
\affiliation{$^4$School of Physical Sciences and Nanotechnology, Yachay Tech University, Urcuqui 100119, Ecuador}

\begin{abstract}
Starting from the rigorous quantum-field-theory formalism we derive a formula for the screened conductivity designed to study the coupling of light with elementary electron excitations and the ensuing electromagnatic modes in two-dimensional (2D) semiconductors. The latter physical quantity consists of three fully separable parts, namely intraband, interband, and ladder conducivities, and is calculated beyond the random phase approximation as well as from first principles.
By using this methodology, we study the optical absorption spectra in 2D black phosphorous, so-called phosphorene, as a function of the concentration of electrons injected into the conduction band. The mechanisms of phosphorene exciton quenching versus doping are studied in detail. It is demonstrated that already small doping levels ($n\sim 10^{12}cm^{-2}$) lead to a radical drop in the exciton binding energy, i.e., from $600$\,meV to  $128$\,meV.  
The screened conductivity is applied to study the collective electromagnetic modes in doped phosphorene. It is shown that the phosphorene transversal exciton hybridizes with free photons to form a exciton-polariton. This phenomenon is experimentally observed only for the case of confined electromagnetic microcavity modes. Finally, we demonstrate that the energy and intensity of anisotropic 2D plasmon-polaritons can be tuned by varying the concentration of injected electrons.  
\end{abstract}

\maketitle

\section{Introduction}
Semiconducting two-dimensional (2D) crystals became very attractive in terms of their very interesting optical and electromagnetic 
properties. Transition metal dichalcogenides (TMD) support tunable \cite{tunedop,tunestrain}
and strong excitons or exciton-polaritons \cite{Nature_Polaritons} in the visible frequncy 
range \cite{TMD-exc1,TMD-exc2}, which can be potentially applied in various optoelectronic devices \cite{TMD-exc-app}. 
Increasing attention has also been given to excitons and trions affected by the dielectric environment \cite{Exp_exMoS2_vs_sub}. Important as well for our understanding of light-matter interaction are the studies of strong hybridization between TMD excitons and dielectric microcavity photons that results in the formation of exciton-polariton modes \cite{ex-pol1,ex-pol2,ex-pol3,ex-pol4,ex-pol5}.
In addition, doped 2D semiconductors can support collective electron excitation modes known as plasmons \cite{Abajo,Politano,Valley-plas}, with promising applications as reported in TMDs/graphene and in gold heterostructures \cite{TMD-plas1,TMD-plas2}. Recently, a new class of 2D materials has emerged that support anisotropic electromagnetic
modes \cite{Anisotrop2D-PRL}. The most famous anisotropic 2D crystal is a single-layer of black phosphorus, also known as phosphorene, which supports tunable 2D hyperbolic plasmon \cite{PhysRevApplied2019}.

The optical properties and dielectric response of phosphorene have been 
systematically investigated\,\cite{Phosp_screen,Abajo,PhysRevApplied2019,pl-sigma,Ph-opt,Ph-pl-pol,Ph-nanoribb,Ph-multil,abinitio-pl1,EELS-MLP,ph-ex1,ph-ex2,ph-ex3,ph-ex4-g079,ph-ex5-Neto-0.87-strain,
ph-ex6-Neto-cited_byEXP4,TDDFT,ph-ex-EXP1,ph-ex-EXP2,ph-ex-EXP4-0.3-SiO2/Si,ph-ex-EXP3}. 
For instance, the intensities and tuning of hyperbolic plasmons in supported or self-standing 
phosphorene were explored by using different models for the optical conductivity, either via tight binding approximation (TBA) fitted to density functional theory (DFT) calculations or via GW methods \cite{Abajo,PhysRevApplied2019,pl-sigma}. Also, optical properties, including optical reflection, transmission, absorption, and plasmon-polaritons in phosphorene, were studied in great detail by means of the TBA optical conductivity tensor \cite{Ph-opt,Ph-pl-pol}. Optical properties of multilayer phosphorene as a function of the number of layers (thickness)\,\cite{ph-ex-EXP3}, doping, and light polarization\,\cite{Ph-multil} were explored. Further, the electron energy loss spectra (EELS) and anisotropic plasmons in phosphorene were studied by means of {\em ab initio} techniques \cite{abinitio-pl1,EELS-MLP}.
Besides the hyperbolic plasmon, phosphorene shows very interesting excitonic effects.   
Sophisticated GW-BSE calculations of the quasi-particle band gap and exciton binding 
energies as a function of strain, polarisation, and dielectric environment were studied in phosphorene \cite{ph-ex1,ph-ex2,ph-ex3,ph-ex4-g079,ph-ex5-Neto-0.87-strain,
ph-ex6-Neto-cited_byEXP4}. Moreover, the excitonic fine structure in monolayer and few-layer 
black phosphorus were studied through reflection and photoluminescence excitation measurements \cite{ph-ex-EXP1}. In Refs.\,\cite{ph-ex-EXP2,ph-ex-EXP4-0.3-SiO2/Si} anisotropic photoluminescence, the quasiparticle band-gap, and the exciton binding energy in phosphorene were studied and compared with theoretical calculations.

These extensive studies have shown that electromagnatic excitations (i.e., plasmon-plaritons and exciton-polaritons) in pristine and doped phosophrene crystals display remarkable optical properties. In this paper we derive a compact formula for the investigation of electromagnatic 
modes in 2D crystals, where the optical conductivity tensor $\sigma_{\mu\nu}(\omega)=
\sigma^{intra}_{\mu\nu}(\omega)+\sigma^{inter}_{\mu\nu}(\omega)+\sigma^{ladd}_{\mu\nu}(\omega)$ is the only input expression and is fully calculated from first principles. 
The first two terms $\sigma^{RPA}=\sigma^{intra}+\sigma^{inter}$ represent the random phase approximation (RPA), while the third term $\sigma^{ladd}$ 
represents the `ladder' contribution to the optical conductivity. In tandem, this becomes the `RPA+ladder' 
approximation. This approach is analogous to the widely used GW-BSE method \cite{BSE1,BSE2,BSE3,BSE4,BSE5,BSE6,BSE7,BSE8}, which is commonly utilized to calculate the quasi-particle and optical properties of various 2D semiconductors \cite{TMD-exc1,TMD-exc2,hBN1,hBN2,hBN3,hBN4,MoS2_1,MoS2_2}, including 
phosphorene \cite{ph-ex1,ph-ex2,ph-ex-EXP2,ph-ex-EXP3}. The `RPA+ladder'  approximation  allows for RPA and ladder terms to be calculated independently, so that the RPA contribution can be calculated 
at the required higher level of accuracy (using many bands and dense ${\bf K}$-points meshes), while the computationally demanding ladder contribution can be calculated by using fewer bands and a coarser ${\bf K}$-point
grid. This could significantly reduce the computational coast while including excitonic effects to a moderate level of accuracy. This is usually not the case in standard BSE calculations where Hartree (RPA) and Fock (ladder) BSE 
kernels form a two-particle hamiltonian (single matrix in energy-momentum space) \cite{hBN1} and must be calculated at 
the same level of accuracy. Also, here the RPA conductivity is further separated into $\sigma^{intra}$ (Drude intraband) and to $\sigma^{inter}$ 
(interband) terms, which facilitates analysis of doped semiconductors. In this paper the `RPA+ladder'  approximation will be applied to study two kinds of electromagnatic modes in doped phosphorene, namely, plasmon- and exciton-polaritons.  

The paper is organized as follows. In Sec.~\ref{theory}, we present 
the derivation of the optical conductivity $\sigma_\mu(\omega)$ in the `RPA+ladder' 
approximation along with the solution of the Dyson equation for the electric field $E_\mu({\bf Q},\omega)$ in 
the vicinity of a 2D crystal.  In Sec.~\ref{Results}, we  demonstrate how the injection of electrons into the phosphorene conduction band (extra electronic 
screening $\Delta W=W_0^{dop}-W_0^0$) influences the principal exciton intensity and binding energy, present results showing the hybridization 
between the exciton and free photons (i.e., formation of exciton-polaritons), and finally show the RPA optical conductivities $\sigma^{RPA}=\sigma^{intra}+\sigma^{inter}$, the effective number of in-plane charge 
carriers $n^{e,h}_\mu$, and the intensities of plasmon-polaritons in doped phosphorene.  
The conclusions are presented in Sec.~\ref{conclu}.
 
\section{Theoretical formulation}
\label{theory} 
The system we explore consists of electrons which move within the effective crystal 
potential and which interact with free photons so that the total Hamiltonian of the system 
can be written as   
\begin{equation}
H=H_{el}+\ H_{ph}+\ H_{el-ph}.
\label{totH}
\end{equation}
Here 
\begin{equation}
H_{el}\ =\ \sum_{n\bf K}E_{n\bf K}c^+_{n\bf K}c_{n\bf K}
\end{equation} 
represents the electrons which move in the effective Kohn-Sham (KS)
potential. The $c^+_{n\bf K}/c_{n\bf K}$ are the creation/annihilation operators of an electron 
in Bloch state $\left|n,{\bf K}\right\rangle$ represented by the wave function $\phi_{n{\bf K}}$ and energy $E_{n{\bf K}}$, where $n$ is the band index and 
${\bf K}=(K_x,K_y)$ parallel wave vector. Analogously,  
\begin{equation}
H_{ph}\ =\ \sum_{\mu\bf q}\hbar\left|{\bf q}\right|c\left\{a^+_{\mu\bf q}a_{\mu\bf q}+\frac{1}{2}\right\}
\label{fre-ph}
\end{equation} 
represents the free photons, where  $a^+_{\mu\bf q}/a_{\mu\bf q}$ are the creation/annihilation operators of a photon with polarization $\mu$, {\bf q} is a three-dimensional (3D) wave vector, and $c$ is the speed of light.
\begin{figure}[t]
\centering
\includegraphics[width=0.4\textwidth]{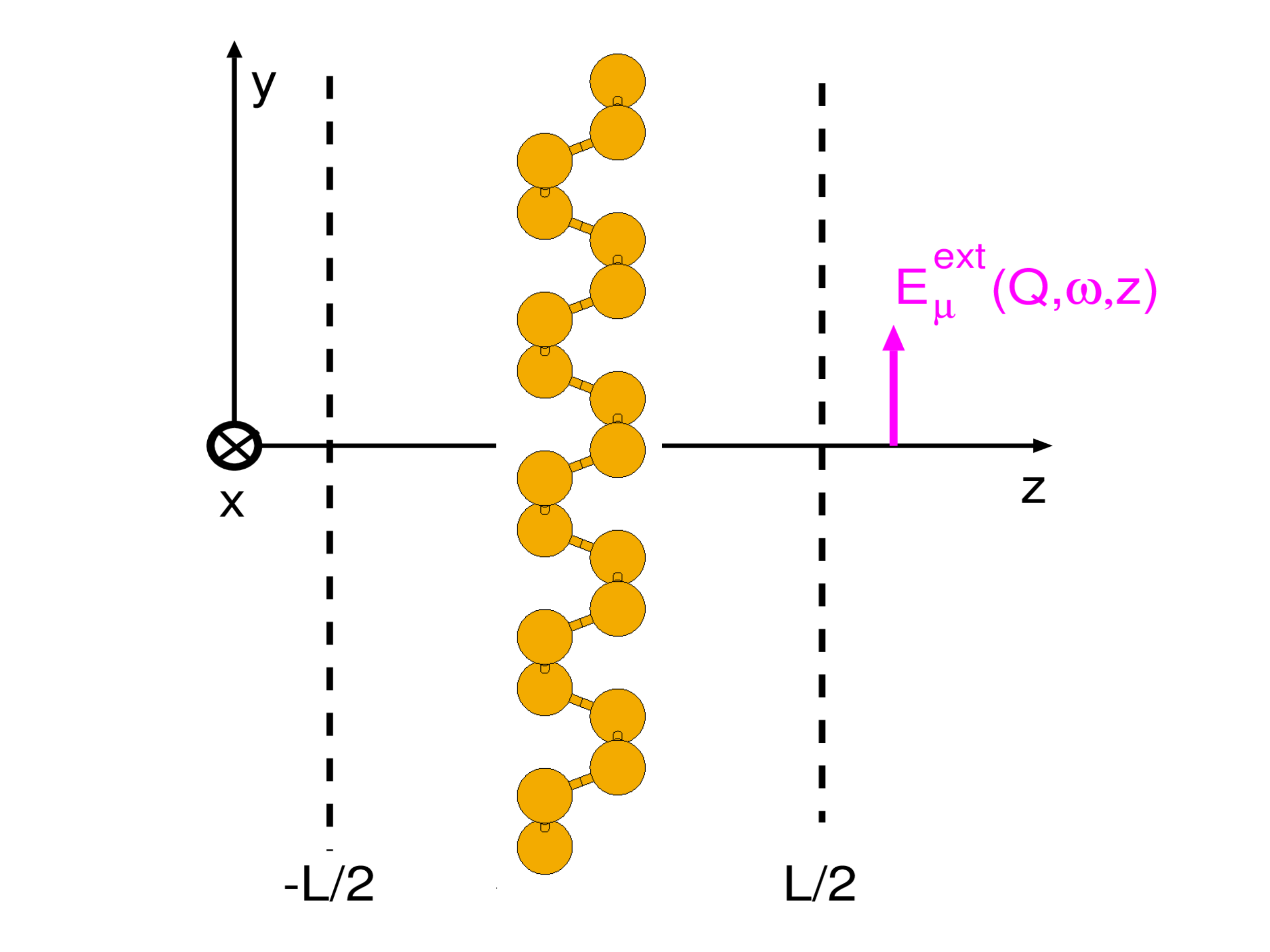}
\caption{Geometry of the system. The 2D crystal (in this case phosphorene) represents one supercell which periodically repeats 
in the perpendicular $z$ direction, where $L$ represents the supercell constant in that direction. 
The volume integration in Eq.\,\ref{Dysons} is restricted within a volume of one supercell $z\in[-L/2,L/2]$ which means that the 
photons can interact solely with the Bloch electrons in the corresponding supercell.}
\label{Fig1}
\end{figure}
In the $\Phi=0$ gauge the part of the Hamiltonian which represents the interaction between electrons and photons can 
be written as\,\cite{Pol,Polariton2016}
\begin{eqnarray}
H_{el-ph}\  = -\frac{1}{c}\int d^3{\bf r}\  {\bf j}\cdot {\bf A}\  +\ \frac{e^2}{2mc^2}\int d^3{\bf r}\ \rho\ {\bf A}^2.  
\label{operatorint}
\end{eqnarray}
Here, ${\bf A}$ is the electromagnetic field or vector potential operator, the fermionic current operator is  
\begin{equation}
{\bf j}\ =\ \frac{e\hbar}{2im}\left\{\Psi^+\nabla\Psi-[\nabla\Psi^+]\Psi\right\},
\label{opferstruj}
\end{equation}
the fermionic density operator is defined as
\[
\rho=\Psi^+\Psi,
\]
and the fermionic field operator is 
\begin{equation}
\psi\left({\bf r}\right)=\sum_{n,{\bf K}}\phi_{n{\bf K}}({\bf r})c_{n{\bf K}}. 
\label{oppo}
\end{equation}
We emphasize here that the spin quantum number $s=\pm 1/2$ will be merged with the bands quantum 
number, i.e. $n\equiv(n,s)$. The time-ordered photon propagator is defined as
\begin{align}
D_{\mu\nu}({\bf r},{\bf r}';t-t')=\frac{i}{\hbar c}
\left\langle\Phi_0\right|T\left\{A_\mu({\bf r},t)A_\nu({\bf r}',t')\right\}\left|\Phi_0\right\rangle,
\label{bozonprop}
\end{align}
where $T$ represents the time ordering operator, $A_\mu(t)\ =\ e^{iHt/\hbar}A_\mu(t=0)e^{-iHt/\hbar}$ is the Heisenberg operator, and $\left|\Phi_0\right\rangle$ 
is a ground state of the total Hamiltoninan in Eq.\,\ref{totH}. After employing the standard perturbation theory method for the bosons Green's functions
\cite{Mahan,Pol,Polariton2016} it can be shown that the photon propagator in Eq.\,\ref{bozonprop} satisfies the Dyson equation 
\begin{eqnarray}
D_{\mu\nu}({\bf r},{\bf r}',\omega)\ =\ D^{0}_{\mu\nu}({\bf r},{\bf r}',\omega)\ +\hspace{2cm} 
\nonumber\\
\sum^3_{\alpha,\beta=1}\int d^2{\brho}_1\int^{L/2}_{-L/2}dz_1\int d^2{\brho}_2\int^{L/2}_{-L/2}dz_2\times\hspace{0cm}
\label{Dysons}\\
D^{0}_{\mu\alpha}({\bf r},{\bf r}_1,\omega)
\Pi_{\alpha\beta}({\bf r}_1,{\bf r}_2,\omega)
D_{\beta\nu}({\bf r}_2,{\bf r}',\omega),\hspace{0cm}
\nonumber
\end{eqnarray}
which is also illustrated with Feynman diagrams in Fig.\,\ref{Fig2}(a). Here the free-photon propagator is  
\begin{align}
D^{0}_{\mu\nu}({\bf r},{\bf r}';t-t')\ =
\frac{i}{\hbar c}\left\langle\Phi^{ph}_0\right|T\left\{A_\mu({\bf r},t)A_\nu({\bf r}',t')\right\}\left|\Phi^{ph}_0\right\rangle,
\label{freephoton}
\end{align} 
where $A_\mu(t)\ =\ e^{iH_{ph}t/\hbar}A_\mu(t=0)e^{-iH_{ph}t/\hbar}$ is the interaction picture operator, and $\left|\Phi^{ph}_0\right\rangle$ is the photonic vacuum 
or ground state of a free-photon Hamiltoninan (see Eq.\,\ref{fre-ph}). 
In this work we restrict ourselves to the `RPA+ladder' approximation  such that  the photon self-energy  
$\Pi$ consists of two terms, i.e., 
\begin{equation}
\Pi_{\alpha\beta}({\bf r},{\bf r}',\omega)=\Pi^{RPA}_{\alpha\beta}({\bf r},{\bf r}',\omega)+\Pi^{ladd}_{\alpha\beta}({\bf r},{\bf r}',\omega),
\end{equation}
where the first `RPA', and the second `ladder' contributions are illustrated by Feynman diagrams in Fig.\,\ref{Fig2}(b) and \ref{Fig2}(c).  
\begin{figure}[b]
\centering
\includegraphics[width=8.5cm,height=6.0cm]{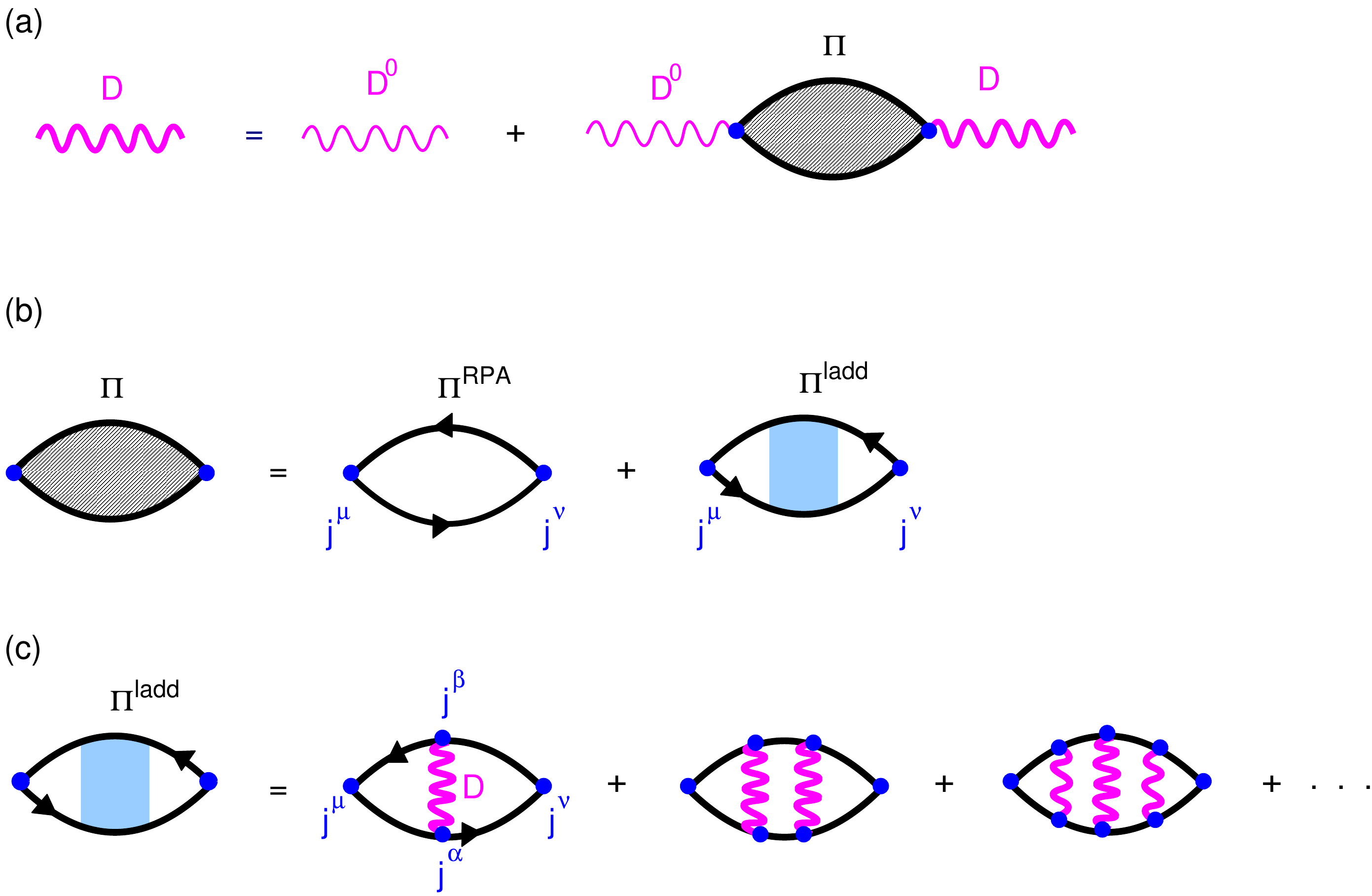}
\caption{(a) Feynman diagrams representing the Dyson equation in Eq.\,\ref{Dysons}. 
(b) Photon self-energy in the `RPA+ladder' approximation. (c) Perturbative expansion of the ladder 
photon self-energy $\Pi^{ladd}$. Blue dots represent the current vertices $j^{\mu}$, black lines 
are single-particle time-ordered Green's functions $G$, and thin and thick magenta wavy lines represent the bare  $D^0_{\mu\nu}$ and the screened $D_{\mu\nu}$  photon 
propagators, respectively.
}
\label{Fig2}
\end{figure}
It should be noted that in Eq.\,\ref{Dysons} the integration is restricted  within the volume of the supercell $z\in[-L/2,L/2]$  (as shown 
in Fig.\,\ref{Fig1}) which cancels the spurious inter-supercell electron-electron interactions. 

In what follows, the methodology used to solve the Dyson equation in Eq.\,\ref{Dysons} is shown with an emphasis on the calculation of the `ladder' photon self-energy, while the detailed derivation of the 
RPA photon self-energy is in Ref.\,\cite{Polariton2016}.  
Considering that the crystal super-lattice is periodic in 3D, all tensors 
can be Fourier expanded as 
\begin{eqnarray}
T_{\alpha\beta}({\bf r},{\bf r}',\omega)=\hspace{6cm}
\nonumber\\
\label{jukmuknew}\\
\frac{1}{L}
\sum_{{\bf G}{\bf G}'}\int\frac{d{\bf Q}}{(2\pi)^2}\ e^{i({\bf Q}+{\bf G}){\bf r}}\ 
e^{-i({\bf Q}+{\bf G}'){\bf r}'}\ T_{\alpha\beta {\bf G}{\bf G}'}({\bf Q},\omega),
\nonumber
\end{eqnarray}
where ${\bf Q}=(Q_x,Q_y)$ is the momentum transfer wave-vector parallel to the $x-y$ plane, 
${\bf G}=({\bf G}_\parallel,G_z)$ are 3D reciprocal lattice vectors and ${\bf r}=(\brho,z)$ is a 3D 
position vector. After using Eq.\,\ref{jukmuknew} the Dyson equation transforms into the matrix equation  
\begin{align}
&D_{\mu\nu,{\bf G}{\bf G}'}({\bf Q},\omega)\ =\ D^{0}_{\mu\nu,{\bf G}{\bf G}'}({\bf Q},\omega)\ +
\nonumber\\
&\sum_{\alpha\beta,{\bf G}_1{\bf G}_2}D^{0}_{\mu\alpha,{\bf G}{\bf G}_1}({\bf Q},\omega)
\Pi_{\alpha\beta,{\bf G}_1{\bf G}_2}({\bf Q},\omega)
D_{\beta\nu,{\bf G}_2{\bf G}'}({\bf Q},\omega).
\label{DysonsG}
\end{align}
The 3D Fourier transform of the free-photon propagator becomes
\begin{eqnarray}
{\bf D}^0_{{\bf G},{\bf G}'}({\bf Q},\omega)=\hspace{6cm}
\label{matD0}\\
\frac{1}{L}\delta_{{\bf G}_{\parallel}{\bf G}_{\parallel}'}
\int^{L/2}_{-L/2}e^{-iG_zz}\ {\bf D}^0({\bf Q}+{\bf G}_{\parallel},\omega,z,z')\ e^{iG'_zz'}dzdz',
\nonumber
\end{eqnarray} 
where the partial Fourier transform of the free-photon 
propagator in the $x-y$ plane is explicitly\,\cite{Pol}  
\begin{eqnarray}
{\bf D}^0({\bf Q},\omega,z,z')=-\frac{4\pi c}{\omega^2}\delta(z-z')
{\bf z}\cdot{\bf z}  
+\hspace{1cm}
\nonumber\\
\frac{2\pi i}{c\beta}
\left\{
{\bf e}_{s}\cdot{\bf e}_{s}+{\bf e}_{p}\cdot{\bf e}_{p}
\right\}e^{i\beta\left|z-z'\right|}.
\label{photQ}
\end{eqnarray} 
Here the unit vectors are adapted to the geometry of the system such that ${\bf e}_{s}={\bf Q}_{0}\times{\bf z}$ and 
${\bf e}_{p}=\frac{c}{\omega}\left[-\beta\ sgn\left(z-z'\right){\bf Q}_{0}+Q{\bf z}\right]$ (where ${\bf Q}_0$ is the unit vector in the ${\bf Q}$ direction) 
represent directions of ${\bf s}$(TE) and ${\bf p}$(TM) polarized fields, respectively. The complex wave vector in the perpendicular ($z$) direction 
is defined as $\beta=\sqrt{\frac{\omega^2}{c^2}-Q^2}$.

\begin{figure}[t]
\centering
\includegraphics[width=8cm,height=8cm]{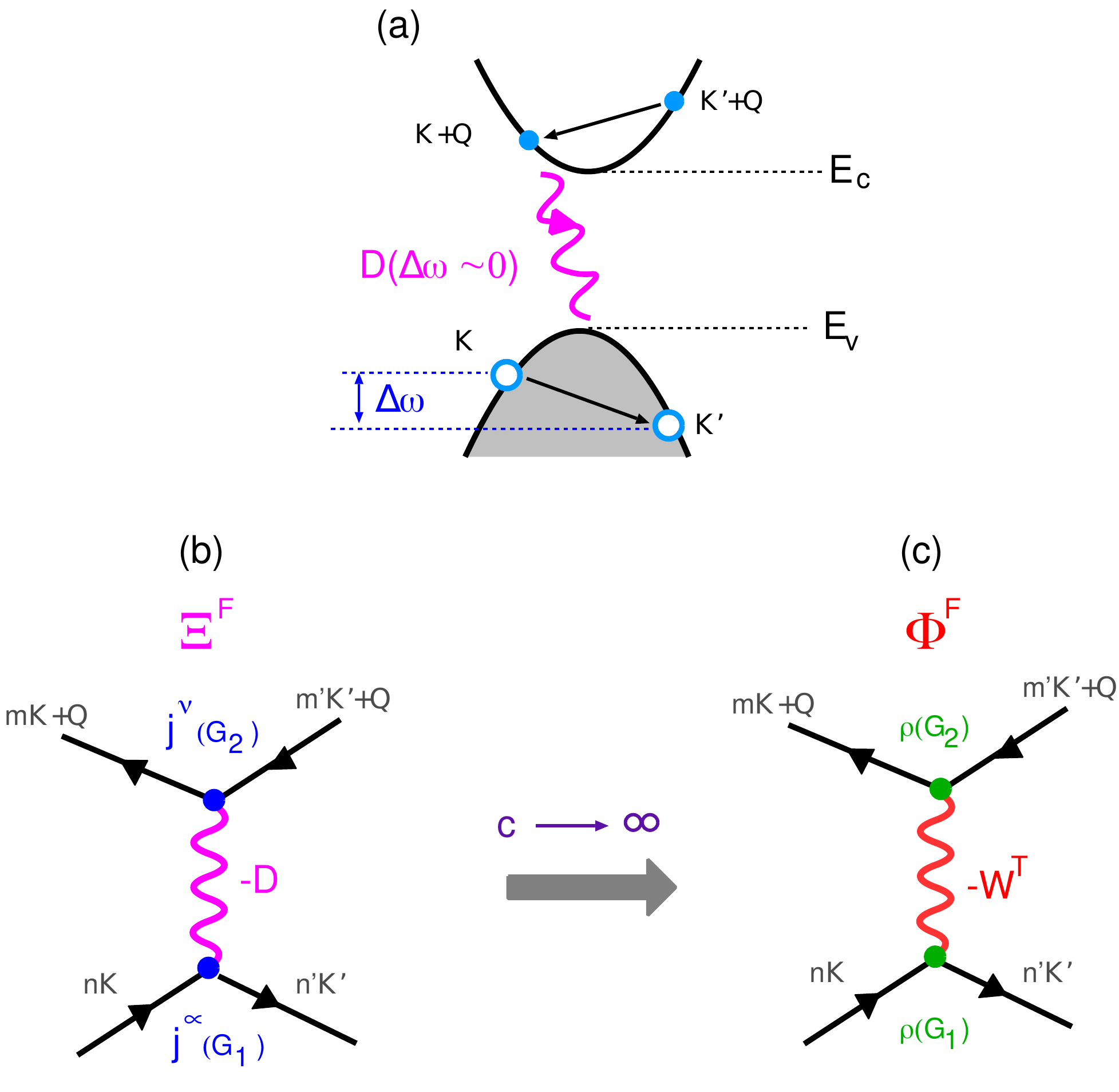}
\caption{(a) The electron hole scattering mediated by the photon propagator $D_{\mu\nu}$. (b) The Feynman diagram 
for the `photonic’ Bethe-Salpeter-Fock kernel in Eq.\,\ref{FockKPh}. (c) In the non-retarded limit ($c\rightarrow\infty$) the `photonic’ Fock kernel $\Xi^{F}$ becomes the standard 
Fock kernel $\Phi^{F}$ in Eq.\,\ref{FockK}, which represents the electron-hole scattering mediated by screened Coulomb interaction $W^T$.}
\label{Fig3}
\end{figure}

\begin{figure}[t]
\centering
\includegraphics[width=8cm,height=5cm]{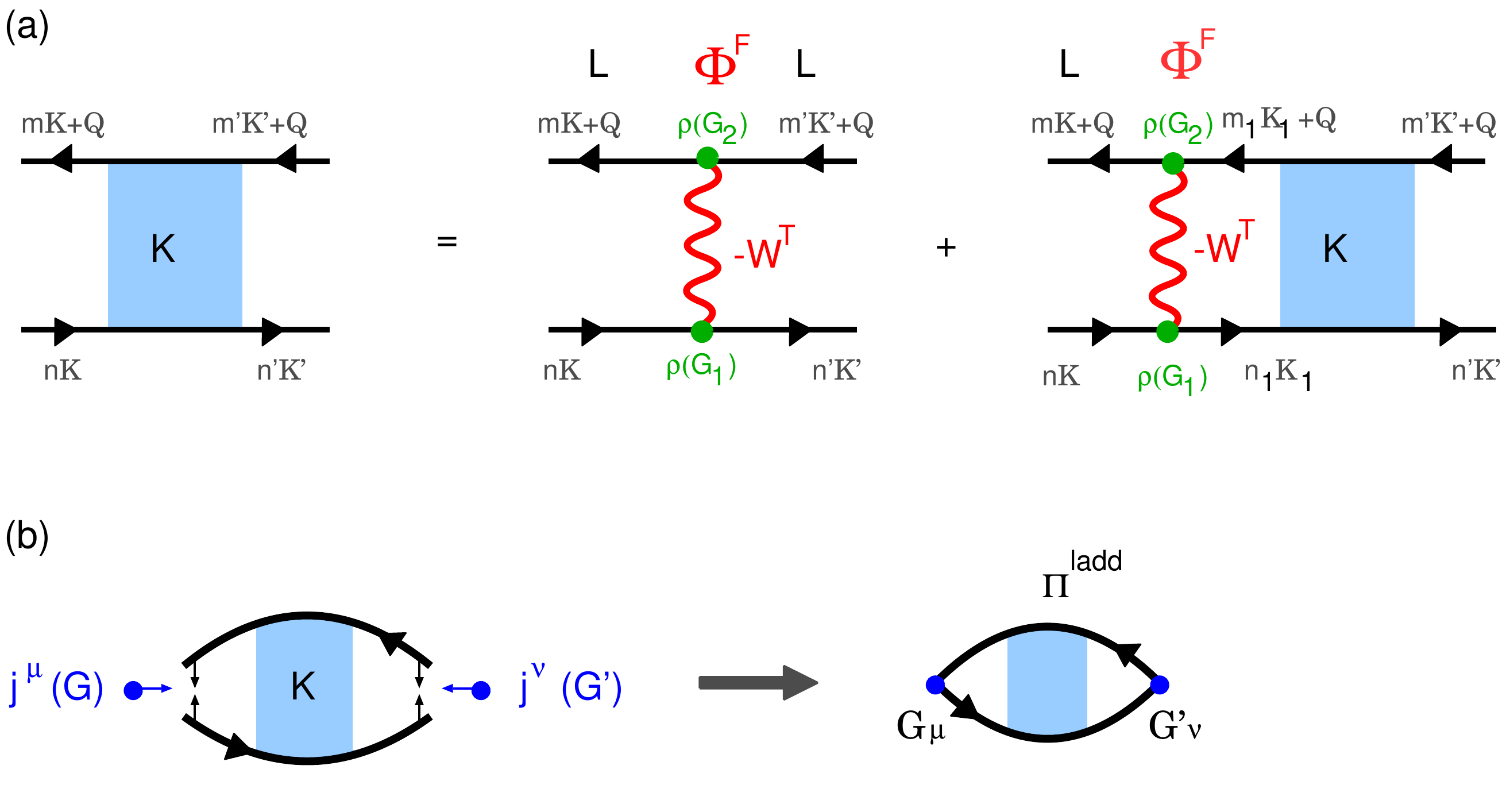}
\caption{The calculation procedure for the ladder photon self-energy $\Pi^{ladd}$. (a) The matrix equation for the ladder 4-point polarizability ${\cal K}$. 
(b) The ladder photon self-energy $\Pi^{ladd}$ is obtained when the fermionic lines in ${\cal K}$ are contracted, multiplied by corresponding 
current vertices $j^{\mu}$ and $j^{\nu}$, and summed over bands and wave vectors,  $\Pi^{ladd}=
-\frac{1}{\Omega c}\sum_{nm{\bf K}}\sum_{n'm'{\bf K}'}\ j^\mu\ {\cal K}^{ladd}\ [j^\nu]^*$.}
\label{Fig4}
\end{figure}

The Fourier transform of the photon self-energy is 
\begin{equation}
\Pi_{\mu\nu,{\bf G}{\bf G}'}({\bf Q},\omega)=
\Pi^{RPA}_{\mu\nu,{\bf G}{\bf G}'}({\bf Q},\omega)+
\Pi^{ladder}_{\mu\nu,{\bf G}{\bf G}'}({\bf Q},\omega),
\label{RPA+ladder}
\end{equation}
where the RPA photon self-energy is explicitly\,\cite{Polariton2016}   
\begin{eqnarray}
&&\Pi^{RPA}_{\mu\nu,{\bf G}{\bf G}'}({\bf Q},\omega)=
\frac{1}{\Omega \mathrm{c}}\sum_{nm{\bf K}}\frac{\hbar\omega}{E_{n{\bf K}}-E_{m{\bf K}+{\bf Q}}}
\nonumber\\
&&\frac{f_{n{\bf K}}-f_{m{\bf K}+{\bf Q}}}
{\hbar\omega+i\eta+E_{n{\bf K}}-E_{m{\bf K}+{\bf Q}}}\nonumber\\
&&j^{\mu}_{n{\bf K},m{\bf K}+{\bf Q}}({\bf G})\ [j^{\nu}_{n{\bf K},m{\bf K}+{\bf Q}}({\bf G}')]^*,
\label{RPAphSE}
\end{eqnarray} 
where $\Omega=S\times L$ is the normalization volume, $S$ is the normalization surface 
and $f_{n{\bf K}}=[e^{(E_{n{\bf K}} - E_F)/kT}+1]^{-1}$ is the Fermi-Dirac distribution function at the temperature $T$. 
The current verices are defined as  
\begin{equation}
j^\alpha_{n{\bf K},m{\bf K}+{\bf Q}}({\bf G})\ =\ 
\int_{\Omega} d{\bf r}e^{-i({\bf Q}+{\bf G}){\bf r}}\ 
j^\alpha_{n{\bf K},m{\bf K}+{\bf Q}}({\bf r}),\hspace{1cm}
\label{kjvcrnk}
\end{equation}
and the current $j^\alpha_{n{\bf K},m{\bf K}+{\bf Q}}({\bf r})$ produced by transition between Bloch states  $\left|n{\bf K}\right\rangle\rightarrow\left|m{\bf K}+{\bf Q}\right\rangle$ is equal to
\begin{eqnarray}
\frac{e\hbar}{2im}\left\{\phi^*_{n{\bf K}}({\bf r})\partial_\alpha\phi_{m{\bf K}+{\bf Q}}({\bf r})-
[\partial_\alpha\phi^*_{n{\bf K}}({\bf r})]\phi_{m{\bf K}+{\bf Q}}({\bf r})\right\}.\hspace{1cm}
\nonumber 
\end{eqnarray}
\noindent
The ladder photon self-energy is 
\begin{align}
&\Pi^{ladd}_{\mu\nu,{\bf G}{\bf G}'}({\bf Q},\omega)
=
-\frac{1}{\Omega c}\sum_{nm{\bf K}}
\sum_{n'm'{\bf K}'}j^{\mu}_{n{\bf K},m{\bf K}+{\bf Q}}({\bf G})\times\nonumber\\
&
{\cal K}^{m{\bf K}+{\bf Q}\leftarrow m'{\bf K}'+{\bf Q}}_{n{\bf K}\rightarrow n'{\bf K}'}(\omega)
[j^{\nu}_{n'{\bf K}',m'{\bf K}'+{\bf Q}}({\bf G}')]^*,
\label{Dkupiuuiug}
\end{align}
where the ladder 4-point polarizability ${\cal K}$ can be obtained by solving the matrix equation in $\left\{{\bf K},n\right\}$-space 
\begin{align}
{\cal K}^{ladd}(\omega)\ =\ 
{\cal L}(\omega)\otimes\Xi^F\otimes{\cal L}(\omega)+{\cal L}(\omega)\otimes\Xi^F\otimes{\cal K}^{ladd}(\omega),
\label{KeqF}
\end{align}  
where matrix multiplication represents summation over the bands 
and wave vectors as $\otimes\ \equiv\ \sum_{nm}\sum_{{\bf K}}$.
Here the time-ordered electron-hole propagator is defined as  
\begin{align}
{\cal L}^{m{\bf K}+{\bf Q}\leftarrow m'{\bf K}'+{\bf Q}}_{n{\bf K}\rightarrow n'{\bf K}'}\ =
\int^{\infty}_{-\infty}\frac{d\omega'}{2\pi i}G_{n{\bf K}}(\omega')G_{m{\bf K}+{\bf Q}}(\omega+\omega').
\label{Lnul}
\end{align}
In the quasi-particle approximation (long lifetime approximation), the time-ordered single-particle propagator 
is defined as 
\begin{equation}
G_{n{\bf K}}(\omega)=
\frac{1-f_{n{\bf K}}}{\omega-E_{n{\bf K}}+i\eta}
+
\frac{f_{n{\bf K}}}{\omega-E_{n{\bf K}}-i\eta},
\label{sulio}
\end{equation}
where the single particle energies $E_{n{\bf K}}$ are
calculated by combining DFT and quasiparticle GW corrections \cite{BSE5}. 
After substituting Eq.\,\ref{sulio} into Eq.\,\ref{Lnul}, the time-ordered electron-hole propagator 
becomes explicitly 
\begin{align}
{\cal L}^{m{\bf K}+{\bf Q}\leftarrow m'{\bf K}'+{\bf Q}}_{n{\bf K}\rightarrow n'{\bf K}'}\ = 
\delta_{nn'}\delta_{mm'}\delta_{{\bf K}{\bf K}'}\times\hspace{2cm}
\nonumber\\
\left\{
\frac{f_{n\bf K}(1-f_{m{\bf K}+{\bf Q}})}
{\omega+E_{n{\bf K}}-E_{m{\bf K}+{\bf Q}}+i\delta}-
\frac{f_{m{\bf K}+{\bf Q}}(1-f_{n{\bf K}})}
{\omega+E_{n{\bf K}}-E_{m{\bf K}+{\bf Q}}-i\delta}\right\}
\label{free-el-hol}
\end{align}
The `photonic' Bethe-Salpeter-Fock kernel is
\begin{align}
&\Xi^{F,m{\bf K}+{\bf Q}\leftarrow m'{\bf K}'+{\bf Q}}_{n{\bf K}\rightarrow n'{\bf K}'}\ = 
-\frac{1}{\Omega c}\sum_{\mu\nu}\sum_{{\bf G}_1{\bf G}_2}\left[j^{\mu}_{n{\bf K},n'{\bf K}'}({\bf G}_1)\right]^* \times
\nonumber\\
&\left[-D^{RPA,\mu\nu}_{{\bf G}_1{\bf G}_2}({\bf K}'-{\bf K},\Delta\omega\approx 0)\right]
j^{\nu}_{m{\bf K}+{\bf Q},m'{\bf K}'+{\bf Q}}({\bf G}_2). 
\label{FockKPh}
\end{align} 
Here, the RPA photon propagator $D^{RPA,\mu\nu}$ is the solution of the Dyson equation (see Eq.\,\ref{DysonsG}) for 
$\Pi_{\mu\nu}=\Pi^{RPA}_{\mu\nu}$, which is explicitly defined in Ref.\,\ref{RPAphSE}.
The photonic Fock-kernel in Eq.\,\ref{FockKPh} represents scattering between excited electrons and 
holes mediated by the photon propagator $D_{\mu\nu}$, as sketched in Fig.\,\ref{Fig3}(a) and in the Feynman diagram in  Fig.\,\ref{Fig3}(b).
Considering that the average electron-hole distance or average exciton radius satisfies 
$r_{ex}\ll c/\Delta\omega$, where $\Delta\omega\approx\hbar^2({\bf K}'^2-{\bf K})/2m_{e,h}$ is electron or hole scattering 
frequency (as sketched in Fig.\,\ref{Fig3}(a)), the interaction between electrons and holes mediated by radiative 
electromagnatic modes (the excitonic  Lamb shift)  is negligible, and in $\Xi^{F}$ we can omit electromagnetic 
retardation effects. This effectively implies that the propagator $D^{\mu\nu}$ can be reduced to the screened Colulomb 
interaction $W^T$, such that the current vertices $j^{\mu}$ become charge vertices $\rho$ and the photon-Fock kernel transform  as  
\begin{eqnarray}
\lim_{c\rightarrow\infty}\Xi^{F,m{\bf K}+{\bf Q}\leftarrow m'{\bf K}'+{\bf Q}}_{n{\bf K}\rightarrow n'{\bf K}'}\ =\ 
\Phi^{F,m{\bf K}+{\bf Q}\leftarrow m'{\bf K}'+{\bf Q}}_{n{\bf K}\rightarrow n'{\bf K}'}(\omega)\ =
\nonumber\\
\frac{1}{\Omega}\sum_{{\bf G}_1{\bf G}_2}
\rho^*_{n{\bf K},n'{\bf K}'}({\bf G}_1)
\left[-W^T_{{\bf G}_1{\bf G}_2}({\bf K}'-{\bf K},\Delta\omega=0)\right] 
\nonumber\\
\rho_{m{\bf K}+{\bf Q},m'{\bf K}'+{\bf Q}}({\bf G}_2),\hspace{2cm}
\label{FockK}
\end{eqnarray}
as shown in Figs.\,\ref{Fig3}(b) and \ref{Fig3}(c). Therefore, the calculation of the ladder photon 
self-energy $\Pi_{\mu\nu}^{ladd}$ consists of  Eqs.\,\ref{Dkupiuuiug}, \ref{KeqF} and \ref{free-el-hol} where instead of the photon 
BSE-Fock kernel Eq.\,\ref{FockKPh} we utilize the ordinary BSE-Fock kernel Eq.\,\ref{FockK}. The calculation procedure of 
the ladder photon self-energy is also illustrated by Feynman diagrams in Fig.\,\ref{Fig4}.    
This computational approach for the `RPA+ladder' photon self-energy is equivalent to solving the Bethe-Salpeter equation within the framework of the widely used 
time-dependent screened Hartree-Fock (TDSHF) approximation\,\cite{BSE1,BSE2,BSE3,BSE4,BSE5,BSE6,BSE7,BSE8}.
The RPA time-ordered screened Coulomb interaction $W^T$, which enters the Fock kernel Eq.\,\ref{FockK}, is obtained by solving the Dyson equation  
\begin{align}
&W^T_{{\bf G}{\bf G}'}({\bf Q},\omega)=
v_{{\bf G}{\bf G}'}({\bf Q})+
\nonumber\\
&\sum_{{\bf G}_1{\bf G}_2}v_{{\bf G}{\bf G}_1}({\bf Q})
\chi^0_{{\bf G}_1{\bf G}_2}({\bf Q},\omega)
W^T_{{\bf G}_2{\bf G}'}({\bf Q}),
\label{WTint}
\end{align}
where the bare Coulomb interaction matrix is  
\begin{align}
v_{{\bf G}_1{\bf G}_2}({\bf Q})=
\delta_{{\bf G}_{1\parallel},{\bf G}_{2\parallel}}
\frac{2\pi}{L|{\bf Q}+{\bf G}_{1\parallel}|}\times\hspace{3cm}
\nonumber\\
\int^{L/2}_{-L/2}dz_1dz_2
e^{-iG_{z1}z_1}
e^{-|{\bf Q}+{\bf G}_{1\parallel}||z_1-z_2|}e^{iG_{z2}z_2}.\hspace{1cm}
\label{bareV}
\end{align}
\newline
The RPA irreducible polarisability is defined as    
\newline
\begin{eqnarray}
\chi^0_{{\bf G}{\bf G}'}({\bf Q},\omega)=
\frac{1}{\Omega}\sum_{nm{\bf K}}\rho_{n{\bf K},m{\bf K}+{\bf Q}}({\bf G})
\rho^*_{n{\bf K},m{\bf K}+{\bf Q}}({\bf G}')
\nonumber\\
\frac{(f_{n\bf K}-f_{m{\bf K}+{\bf Q}})}{\hbar\omega+E_{n\bf K}-E_{m{\bf K}+{\bf Q}}+i\delta sgn(E_{m{\bf K}+{\bf Q}}-E_{n\bf K})},\hspace{1cm} 
\label{chi0}
\end{eqnarray}
\newline
where the charge vertices are 
\newline
\begin{equation}
\rho_{n{\bf K},m{\bf K}+{\bf Q}}({\bf G})=
\int_{\Omega}d{\bf r}\  
\phi^*_{n{\bf K}}({\bf r})
e^{-i({\bf Q}+{\bf G}){\bf r}}
\phi_{m{\bf K}+{\bf Q}}({\bf r}).
\label{dvdnuu}
\end{equation}

\subsection{Optical limit $Q\approx 0$.} 
Here we shall explore electromagnetic modes in 2D crystals in the frequency range going from the terahertz (THz) up to ultraviolet (UV) values, i.e., $\hbar\omega\lesssim 4$\,eV. The  corresponding wavelength is 
then much larger than the 2D crystal thickness and the parallel unit cell size, i.e., $\lambda=\frac{2\pi c}{\omega}\gg L,a$. Therefore, the electromagnetic field variations on the scale of one unit cell are irrelevant and the crystal-local-field effects (CLFE) can be freely excluded from consideration. 
Setting ${\bf G}={\bf G}'=0$ into Dyson's equation yields
\begin{eqnarray}
D_{\mu\nu}({\bf Q},\omega)\ =\ D^{0}_{\mu\nu}({\bf Q},\omega)\ + \hspace{3cm}
\label{DysonsG0}\\
\sum^3_{\alpha,\beta=1}
D^{0}_{\mu\alpha}({\bf Q},\omega)
\Pi_{\alpha\beta}({\bf Q},\omega)
D_{\beta\nu}({\bf Q},\omega).
\nonumber
\end{eqnarray}
Here we have introduced the term $\Pi_{\mu\nu}({\bf Q},\omega)=\Pi_{\mu\nu,{\bf G}=0{\bf G}'=0}({\bf Q},\omega)$, 
$D^{0}_{\mu\nu}({\bf Q},\omega)=D^0_{\mu\nu,{\bf G}=0{\bf G}'=0}({\bf Q},\omega)$.
Dyson's equation (see Eq.\,\ref{DysonsG0}) is expressed in terms of abstract tensors $D$ and $\Pi$, but 
we shall rewrite it in terms of measurable quantities, i.e.,  the electric field $E_\mu$ and conductivity $\sigma_{\mu\nu}$. The screened vector potential produced by an external current  ${\bf j}^{ext}$ 
is defined as 
\begin{equation}
A_\mu({\bf Q},\omega)=\sum^{3}_{\nu=1}D_{\mu\nu}({\bf Q},\omega)j_\nu^{ext}({\bf Q},\omega), 
\label{VecPot}
\end{equation}
while the bare vector potential is analogously defined as ${\bf A}^{ext}=\hat{D}^0{\bf j}^{ext}$. In the $\Phi=0$ gauge 
the connection between the vector potential and the electric field is  
\begin{equation}
E_{\mu}({\bf Q},\omega)=
\frac{i\omega}{c}A_{\mu}({\bf Q},\omega).
\label{elfiT}
\end{equation} 
Moreover, combining formulas  ${\bf j}^{ind}=\hat{\Pi}{\bf A}$, ${\bf j}^{ind}=\hat{\sigma}{\bf E}$, and Eq.\,\ref{elfiT}, 
one obtain the connection between the photon self-energy and the conductivity tensor  
\begin{equation}
\sigma_{\mu\nu}(\omega)=\frac{c}{i\omega}\Pi_{\mu\nu}({\bf Q},\omega).
\label{optcond}
\end{equation}
After substitution of Eqs.\,\ref{VecPot}--\ref{optcond} into Eq.\,\ref{DysonsG0} we obtain the Dyson equation for the screened electric field 
\begin{eqnarray}
E_{\mu}({\bf Q},\omega)\ =\ E^{ext}_{\mu}({\bf Q},\omega)\ + \hspace{3cm}
\label{DysonsE}\\
\sum^3_{\alpha,\beta=1}
\Gamma_{\mu\alpha}({\bf Q},\omega)
\sigma_{\alpha\beta}({\bf Q},\omega)
E_{\beta}({\bf Q},\omega),
\nonumber
\end{eqnarray}
where we introduce the propagator of the free electric field 
\begin{equation} 
\Gamma_{\mu\nu}({\bf Q},\omega)=\frac{i\omega}{c}D^{0}_{\mu\nu}({\bf Q},\omega).
\label{Gamma}
\end{equation}
The formal solution of Eq.\,\ref{DysonsE} is therefore 
\begin{equation}
{\bf E}({\bf Q},\omega)\ =\hat{\epsilon}^{-1}({\bf Q},\omega){\bf E}^{ext}({\bf Q},\omega),
\label{el-field}
\end{equation}
where the dielectric tensor is defined as
\begin{equation}
\epsilon_{\mu\nu}({\bf Q},\omega)=\delta_{\mu\nu}-\sum^3_{\alpha=1}\Gamma_{\mu\alpha}({\bf Q},\omega)\sigma_{\alpha\nu}({\bf Q},\omega).
\label{dieltensor}
\end{equation}
In the THz- and UV- regions ($\omega\lesssim 4$\,eV) and for the parallel wave vector $QL\ll 1$, the complex perpendicular wave 
vector goes as $\beta L \rightarrow 0 $. Using Eq.\,\ref{Gamma} and expressions defined in Eqs.\,\ref{matD0}--\ref{photQ}, 
the free electric field propagator can be approximated as
\begin{align}
{\bf\Gamma}({\bf Q},\omega)=
-\frac{2\pi\beta L}{\omega}{\bf Q}_{0}\cdot{\bf Q}_{0}
-\frac{2\pi\omega L}{\beta c^2}{\bf e}_{s}\cdot{\bf e}_{s}-
\nonumber\\
\left[\frac{4i\pi}{\omega}+\frac{2\pi Q^2L}{\beta\omega}\right]{\bf z}\cdot{\bf z}.\hspace{0.5cm}  
\label{GammaM}
\end{align}
In the  optical limit the conductivity tensor can be approximated as a diagonal matrix 
\begin{equation}
\sigma_{\mu\nu}({\bf Q}\approx 0,\omega)\approx\sigma_{\mu}(\omega)\delta_{\mu\nu},
\label{MatrixS}
\end{equation}
where, following Eq.\ref{optcond}, the optical conductivity  is given by
\begin{equation}
\sigma_\mu(\omega)=\frac{c}{i\omega}\Pi_{\mu\mu,{\bf G}=0{\bf G}'=0}({\bf Q}=0,\omega).
\label{defOptcond}
\end{equation}
According to Eq.\,\ref{RPA+ladder} the optical conductivity can also be 
separated into the `RPA' and `ladder' contributions. Moreover, because here we study doped semiconductors,
it is useful to additionally separate the RPA term into intra- and inter-band contributions  
such that total conductivity can be written as 
\begin{equation}
\sigma_{\mu}(\omega)=\sigma^{\mathrm{intra}}_{\mu}(\omega)\ +\ \sigma^{\mathrm{inter}}_{\mu}(\omega)\ +\ \sigma^{ladd}_{\mu}(\omega).
\end{equation}
After using Eqs.\,\ref{defOptcond} and \ref{RPAphSE} the intraband ($n=m$) RPA  optical 
conductivity is defined as
\begin{equation}
\sigma^{\mathrm{intra}}_{\mu}(\omega)=i\frac{e^2}{m}\ \frac{n_\mu}{\omega+i\eta_ {intra}},
\label{Drudecond}
\end{equation}
where the effective number of charge carriers is 
\begin{equation}
n_\mu\ =\ -\frac{m}{\Omega e^2}\ \sum_n\sum_{{\bf K}\in 1.SBZ}\ \frac{\partial f_{n{\bf K}}}{\partial E_{n{\bf K}}}
\ \left|j^\mu_{n{\bf K},n{\bf K}}({\bf G}=0)\right|^2.
\label{effectivencdc}
\end{equation}
The interband ($n\ne m$) RPA optical conductivity is given by
\begin{eqnarray}
\sigma^{\mathrm{inter}}_{\mu}(\omega)\ =\ -i\frac{\hbar}{\Omega}\ 
\sum_{n\ne m}\sum_{{\bf K}\in 1.SBZ}\ 
\frac{f_{n{\bf K}}-f_{m{\bf K}}}{E_{n{\bf K}}-E_{m{\bf K}}}\times\hspace{1cm}
\nonumber\\
\frac{\left|j^\mu_{n{\bf K},m{\bf K}}({\bf G}=0)\right|^2}{\hbar\omega+E_{n{\bf K}}-E_{m{\bf K}}+i\eta_{inter}}.\hspace{1cm}
\label{Intracond}
\end{eqnarray}
Following the definition given in Eq.\,\ref{defOptcond}, the ladder optical conductivity is explicitly given by
\begin{equation}
\sigma^{ladd}_{\mu}(\omega)=\frac{c}{i\omega}\Pi^{ladd}_{\mu\mu,{\bf G}=0{\bf G}'=0}({\bf Q}=0,\omega),
\label{laddopc}
\end{equation}
where the calculation of the ladder self-energy is described by Eqs.\,\ref{Dkupiuuiug}--\ref{dvdnuu}. 
We stated previously that neglecting the CLFE in the photon self-energy $\Pi$ is fully justified. However, while 
calculating the ladder contribution $\Pi^{ladd}$, one should be careful when neglecting CLFE in the Fock kernel (see Eq.\,\ref{FockK}). Short range ${\bf K}'-{\bf K}\sim 2\pi/a,2\pi/L$  electron-electron (or hole-hole) scattering processes can occur, thereby making exclusion of CLFE in the Coulomb interaction $W^T({\bf K}'-{\bf K})$ not completely justified. Nevertheless, as we shall demonstrate in Sec.\,\ref{Results}, since the main contribution to the exciton binding 
energy  comes from the scattering proceses with ${\bf K}'-{\bf K}<2\pi/a,2\pi/L$, disregarding the CLFE in the Fock-kernel Eq.\,\ref{FockK} still serves as a satisfactory approximation. 
In this approximation, Dyson's equation Eq.\,\ref{WTint} becomes a scalar equation, where the solution is
\begin{equation}
W^T({\bf Q},\omega)=v_Q/\epsilon({\bf Q},\omega),
\label{screenW}
\end{equation}
with the longitudinal dielectric function given by
\begin{equation}
\epsilon({\bf Q},\omega)=1-v_Q\chi^0({\bf Q},\omega).
\end{equation}
Using Eq.\,\ref{bareV}, the bare Coulomb interaction is
\begin{equation}
v_Q=v_{{\bf G}=0{\bf G}'=0}({\bf Q})=
\frac{4\pi}{Q^2}
\ \frac{QL+e^{-QL}-1}
{QL},
\end{equation}
\newline
and by following Eq.\,\ref{chi0} the RPA irreducible 
polarizability becomes
\begin{eqnarray}
\chi^0({\bf Q},\omega)=\chi^0_{{\bf G}=0{\bf G}'=0}({\bf Q},\omega)=
\frac{1}{\Omega}\sum_{nm{\bf K}}(f_{n\bf K}-f_{m{\bf K}+{\bf Q}})\times
\nonumber\\
\frac{\left|\rho_{n{\bf K},m{\bf K}+{\bf Q}}({\bf G}=0)\right|^2}
{\hbar\omega+E_{n\bf K}-E_{m{\bf K}+{\bf Q}}+i\delta sgn(E_{m{\bf K}+{\bf Q}}-E_{n\bf K})}.\hspace{1cm} 
\label{chi02D}
\end{eqnarray}
\subsubsection{Spectra of electromagnatic modes} 
In anisotropic 2D crystals (such as phosphorene) the intensity of the electromagnetic 
modes depends on the direction of its propagation ${\bf Q}_{0}$, such that ${\bf Q}_{0}$ can not be chosen 
arbitrarilyy and the electric field propagator matrix in  Eq.\,\ref{GammaM} remains generally nondiagonal. 
However, here we shall restrict our consideration to the electromagnetic modes 
which propagate in ${\bf Q}_{0}\ ={\bf x}$ and ${\bf Q}_{0}\ ={\bf y}$ directions, i.e., along the phosphorene ${\bf a}_1$ and ${\bf a}_2$ crystal axes, respectively. 
For example, if the electromagnetic mode propagates in the ${\bf Q}_{0}\ =\ {\bf x}$ direction, the free electric field propagator in Eq.\,\ref{GammaM} becomes the diagonal 
matrix 
\begin{equation}
\Gamma_{\mu\nu}(Q{\bf x},\omega)=\Gamma_\mu\delta_{\mu\nu},
\label{MatrixG}
\end{equation} 
where
$\Gamma_x=-\frac{2\pi \beta L}{\omega}$, $\Gamma_y=-\frac{2\pi\omega L}{\beta c^2}$ and $\Gamma_z=-\frac{4i\pi}{\omega}-
\frac{2\pi Q^2L}{\beta\omega}$. 
After combining Eqs.\,\ref{dieltensor},   \ref{MatrixS}, and \ref{MatrixG}, the dielectric tensor can be expressed explicitly as
\begin{equation}
\epsilon_{\mu\mu}(Q{\bf x},\omega)=1-\Gamma_{\mu}\sigma_\mu(\omega).
\label{epsilonT}
\end{equation}
The electromagnetic mode propagation in the ${\bf Q}_{0}\ =\ {\bf y}$ 
direction is given by making the substitution $\Gamma_{x}\leftrightarrow\Gamma_{y}$.
Finally, by following Eq.\ref{el-field} the screened electric field is 
\begin{equation}
E_\mu({\bf Q},\omega)\ =E_\mu^{ext}({\bf Q},\omega)/\epsilon_{\mu\mu}({\bf Q},\omega).
\label{el-ind}
\end{equation}
The induced current is defined as a response function of the screened field via
\begin{equation}
j^{ind}_\mu({\bf Q},\omega)\ =\sigma_\mu(\omega)E_\mu({\bf Q},\omega).
\label{ind-cur}
\end{equation}
Substitution of Eq.\,\ref{el-ind} into the above equation yields the induced current as a response to the external field as   
\begin{equation}
j^{ind}_\mu({\bf Q},\omega)\ =\sigma^{scr}_\mu({\bf Q},\omega)E^{ext}_\mu({\bf Q},\omega),
\label{ind-cur}
\end{equation}
where we introduce the screened conductivity 
\begin{equation}
\sigma^{scr}_\mu({\bf Q},\omega)\ =\sigma_\mu(\omega)/\epsilon_{\mu\mu}({\bf Q},\omega).
\label{ind-cur}
\end{equation}
The real part of the optical conductivity Re$\left[\sigma_\mu(\omega)\right]$
gives us information about the intensity of optically active interband transitions and 
excitons in the system. On the other hand, the real part of the screened conductivity Re$\left[\sigma^{scr}_\mu({\bf Q},\omega)\right]$ 
gives information about the collective electronic modes and hybridizations between electronic modes and photons,  
such as plasmon-polaritons and exciton-polaritons. Therefore, the present formulation enables us to explore 
a wide class of electromagnetic modes, such as evanescent $\omega<Qc$, radiative $\omega>Qc$, 
transverse s(TE) $\sigma^{scr}_{x(y)}[Q{\bf y}({\bf x}),\omega]$, or longitudinal p(TM) $\sigma^{scr}_{x(y)}[Q{\bf x}({\bf y}),\omega]$ 
single particle and collective electromagnetic modes.

\subsubsection{Clarification of Terminology}

In order to facilitate the understanding of the text, we shall first clarify some of the labels and definitions that are often used 
below. The screened Coulomb interaction in pristine phosphorene obtained from the
KS wave function and energies  Eqs.\,\ref{screenW}--\ref{chi02D} will be denoted as $W^{0}_0$. 
The same screened interaction but in doped phosphorene will be denoted as $W^{dop}_0$.
Similarly, Green's functions Eq.\,\ref{sulio} constructed from the pristine or doped phosphorene KS wave function and 
energies will be denoted as $G^0_0$ or $G^{dop}_0$, respectively.    
In pristine semiconductors, the energetic onset for the creation of non-interacting (RPA) electron-hole 
pairs is the band gap energy $E_g=E_C-E_V$, where $E_V$ is the top of the valence band and $E_C$ is the bottom of the conductive 
band, also denoted in the phosphorene band structure shown in Fig.\,\ref{Fig5}. In 
semiconductors doped by electrons ($n>0$), and for reasonably small temperatures ($T<300K$), the value $E_g$ should be, due to Pauli blocking, corrected by an amount 
$2(E_F-E_C)$ such that the onset for the RPA electron-hole pair creation becomes $E_g+2(E_F-E_C)$. Consequently, the exciton binding energy is defined as 
\begin{equation}
\Delta=E_g+2\alpha(E_F-E_C)-\hbar\omega_{ex},
\label{ExBE}
\end{equation}
where $\alpha=0$ and $1$ for pristine ($n=0$)  and doped ($n>0$) semiconductors, respectively, and $\hbar\omega_{ex}$ is the exciton energy. 
The abbreviation RPA$(G^i_0)$, where  $i=0$ or $dop$, will denote the RPA method in which the Green's functions 
G$^i_0$ is inserted. Also, with BSE$(G^i_0,W^i_0)$ ($i=0,dop$) we denote the `RPA+ladder' method, where the Green's functions 
G$^i_0$ are used, and the screened Coulomb interaction W$^i_0$ enters the BSE-Fock kernel Eq.\,\ref{FockK}.  
In all cases it is understood that the Green's function $G^i_0$ is constructed from GW$_0$  energies 
$E^i_{n{\bf K}}$ ($i=0,dop$). The extra screening that comes from the doping is labelled as $\Delta W=W^{dop}_0-W^{0}_0$.

\subsection{Computational details}
\label{Comp}
In the first stage of calculations, we determine the phosphorene KS wave functions $\phi_{n{\bf K}}$ and energies 
$E_{n{\bf K}}$ using a plane-wave self-consistent field DFT code (PWSCF) within the QUANTUM ESPRESSO (QE) package \cite{QE}. 
The core-electron interactions were approximated by norm-conserving pseudopotentials \cite{normcon}, and the exchange-correlation (XC) potential by the Perdew-Burke-Ernzerhof (PBE) generalized gradient approximation (GGA)\,\cite{PBE}. To calculate the ground state electronic density we have used a $26\times37\times1$ Monkhorst-Pack K-point mesh\,\cite{MPmesh} of the first Brillouin zone 
(BZ) and for the plane-wave cut-off energy we have choosen $50$\,Ry. We have used the orthorhombic Bravais lattice where the unit cell lattice constants of $a=4.631\,{\rm \AA}$ and $b=3.3062\,{\rm \AA}$, while the separation between phoshporene layers is given by $L=17.11\,{\rm \AA}$. The doped phosphorene was simulated such that extra electrons were injected ($n>0$) or extracted ($n<0$) from the  unit 
cell and  the compensating jellium background was inserted to neutralize the unit cell. The electronic and atomic relaxation were provided for 
each doping concentration $n$ until a maximum force below $0.001$\,Ry/a.u. was obtained. The RPA optical conductivity Eqs.\,\ref{Drudecond}--\ref{Intracond}
and screened Coulomb interaction in Eqs.\,\ref{screenW}--\ref{chi02D} were calculated by using a $109\times151\times1$ K-point mesh, and the band summations were performed over $50$ bands. The dimension of $\left\{{\bf K},n\right\}$-space used in the calculation 
of the BSE-Fock kernel Eq.\,\ref{FockK}, the 4-point polarisability matrix Eq.\,\ref{KeqF}, the ladder photon self-energy Eq.\,\ref{Dkupiuuiug}, and the ladder optical 
conductivity Eq.\,\ref{laddopc} consists of $53\times75\times1$ Monkhorst-Pack K-points and two (one valence and one conduction) bands. 
The CLFE are not included in the calculation. The DFT calculations underestimate the semiconducting band-gap which then influences the total excitation spectra as well as the exciton 
energy $\hbar\omega_{ex}$. In order to overcome this issue, the energies $E_{n{\bf K}}$ used to calculate the `RPA+ladder' conductivities (for each doping concentration $n$) were 
obtained by means of the GW quasiparticle approximation as implemented within the real space projector augmented wavefunction (PAW) code \textsc{gpaw}\,\cite{GPAW,GPAWRev}. The corresponding ground state parameters and crystal structures follow those outlined for the QE calculations. We have used the $20\times 30\times 1$ K-grid. 100 bands for the GW calculation were used, and the energy cutoff for the local field effect vectors is 80\,eV. The self-consistent GW$_0$ method with $n=3$ steps was used, where energies in the Green's functions are iterated.

In order to check the accuracy of the here introduced 'RPA+ladder' appoximation the results for exciton spectra, exciton energies $\hbar\omega$ and binding energies 
$\Delta$ are compared with results obtained by means of GPAW, where optical properties with excitonic effects included can be obtained by solving the BSE effective two-particle Hamiltonian. In order to solve the BSE within the GPAW code we have used the $53\times75\times1$ K-grid, 10\,eV energy cutoff for the CLFE, and 4 (two valence and two conduction) bands. The broadening parameter was set to 0.05\,eV.

\section{Results} 
\label{Results}
Here we shall first present the results for the optical conductivity Re\,$\sigma_x(\omega)$ in doped 
phosphorene for various doping concentrations $n$, as $y$ polarised light yields no excitonic response\,\cite{ph-ex-EXP2}.  Then we shall 
present the results for the screened conductivity Re\,$\sigma^{scr}_x({\bf Q},\omega)$ 
for different wave vector directions, i.e., $Q_{y}$ and $Q_{x}$, where transverse exciton-polaritons and longitudinal excitons are found, respectively. 
Finally, we shall present results for Re\,$\sigma^{scr}_{x(y)}(Q_{x(y)},\omega)$ in the THz frequency 
region, where longitudinal plasmon-polaritons are formed.

\subsection{Optical conductivity in doped phosphorene} 
\label{OPTc}

Fig.\,\ref{Fig6}(a) shows plots for the RPA($G^0_0$) (black) and BSE($G^0_0,W^0_0$) (magenta) optical conductivities in pristine 
phosphorene. The GW band gap in pristine phosphorene is $E_g=2.05$\,eV, and the RPA conductivity shows an onset for electron-hole creation at 
the same energy. The  BSE($G^0_0,W^0_0$) conductivity shows a strong exciton at $\hbar\omega_{ex}=1.45$\,eV whose binding energy, according 
to Eq.\,\ref{ExBE}, is $\Delta=600$\,meV. This value underestimates the theoretical 
results $\Delta\sim 0.6-0.8$eV reported in Refs.\,\cite{ph-ex1,ph-ex2,ph-ex3,ph-ex4-g079,ph-ex5-Neto-0.87-strain,
ph-ex6-Neto-cited_byEXP4} as well as experimental reults $\Delta\sim 0.9$eV reported in Refs.\,\cite{ph-ex-EXP1,ph-ex-EXP2,ph-ex-EXP3}. However, the exciton binding energy is not easy to determine experimentally because (1) the band-gap $E_g$ is difficult to measure 
accurately, (2) even very small substrate-induced doping of the phosphorene conducting/valence bands causes a screening shift 
$\Delta W$ which can significantly change the exciton binding energy, and (3) the substrate Coulomb screening also influences the exciton 
binding energy. All these may lead to the disparate results seen, such that for example in Ref.\,\cite{ph-ex-EXP2} the binding energy is estimated to be $\Delta=0.9$\,eV, and 
in Refs.\cite{ph-ex-EXP4-0.3-SiO2/Si,ph-ex6-Neto-cited_byEXP4}, where the phosphorene is deposited on the SiO$_2$/Si substrate, 
it is estimated as $\Delta=0.3$\,eV. Still, in order to ensure that the results obtained using the `RPA+ladder' approach are satisfactorily accurate, the 
green  dashed line in Fig.\,\ref{Fig6}(a) shows the result obtained by solving GW-BSE using the \textsc{gpaw} package. 
Besides very good qualitative agreement between the two spectra, the \textsc{gpaw} exciton energy is $\hbar\omega_{ex}=1.51$\,eV and the exciton binding 
energy is $\Delta=540$meV, both of which are in satisfactorily good agreement with the results of our calculations.
Also, while it is often assumed that the exciton energy $\hbar\omega$ does not depend on the substrate 
screening\,\cite{Exp_exMoS2_vs_sub}, this is not always the case. Below, we shall decompose different mechanisms affecting the final exciton spectra when the phosphorene is doped by electrons.      

\begin{figure}[t]
\centering
\includegraphics[width=8cm,height=5cm]{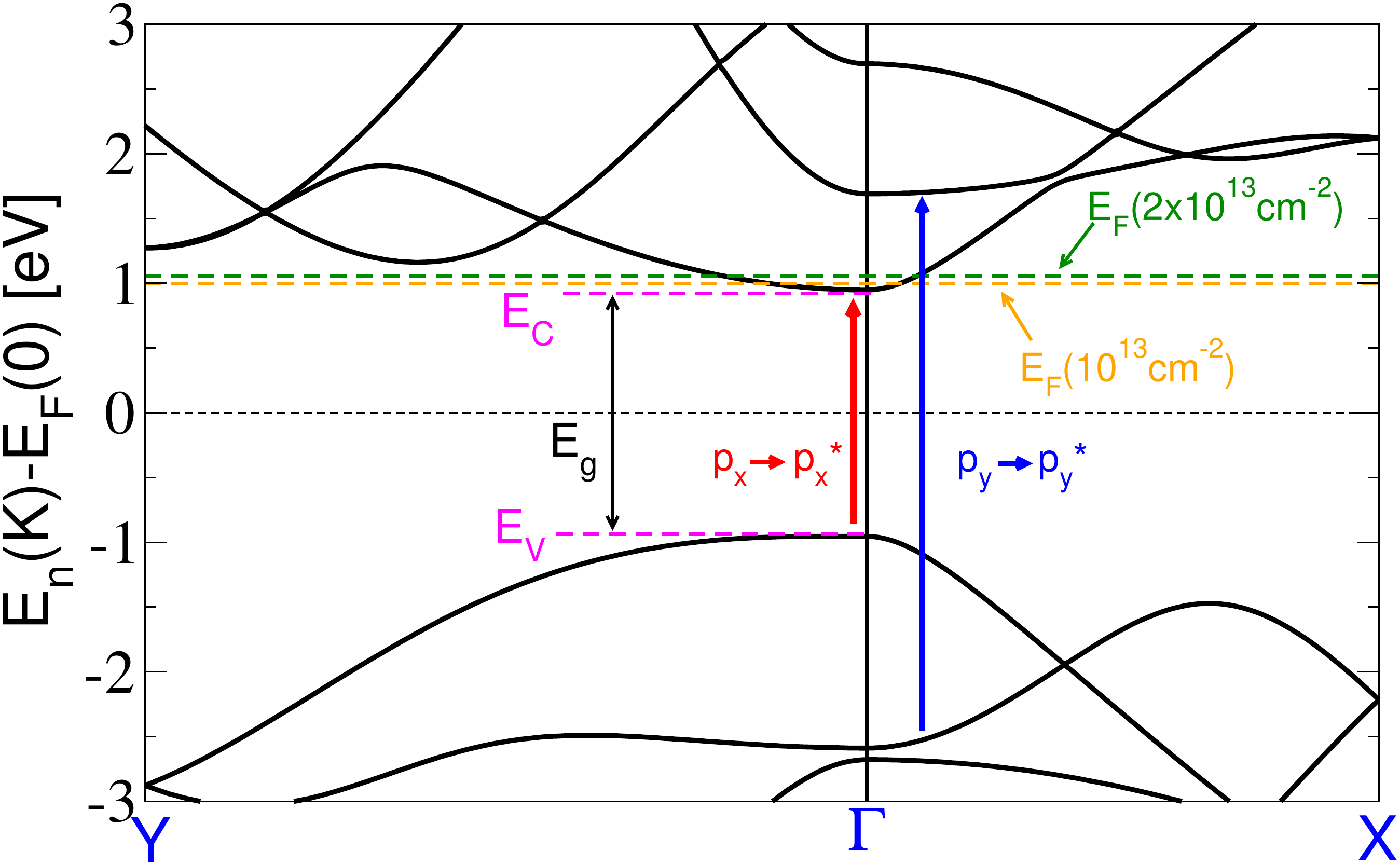}
\caption{The phosphorene band structure. Yellow and green dashed lines denote the Fermi energy in doped phosphorene at $T=284K$ when $n=10^{13}$cm$^{-2}$ and 
$n=2\times10^{13}$cm$^{-2}$, respectively.  The pristine Fermi energy is set to zero. The Fermi energies which correspond
to doped phosphorene with $n=10^{13}$cm$^{-2}$ and $n=2\times10^{13}$cm$^{-2}$ are  $E_F-E_C=52$\,meV and $108$\,meV, respectively.}
\label{Fig5}
\end{figure}
\begin{figure}[t]
\centering
\includegraphics[width=8cm,height=11cm]{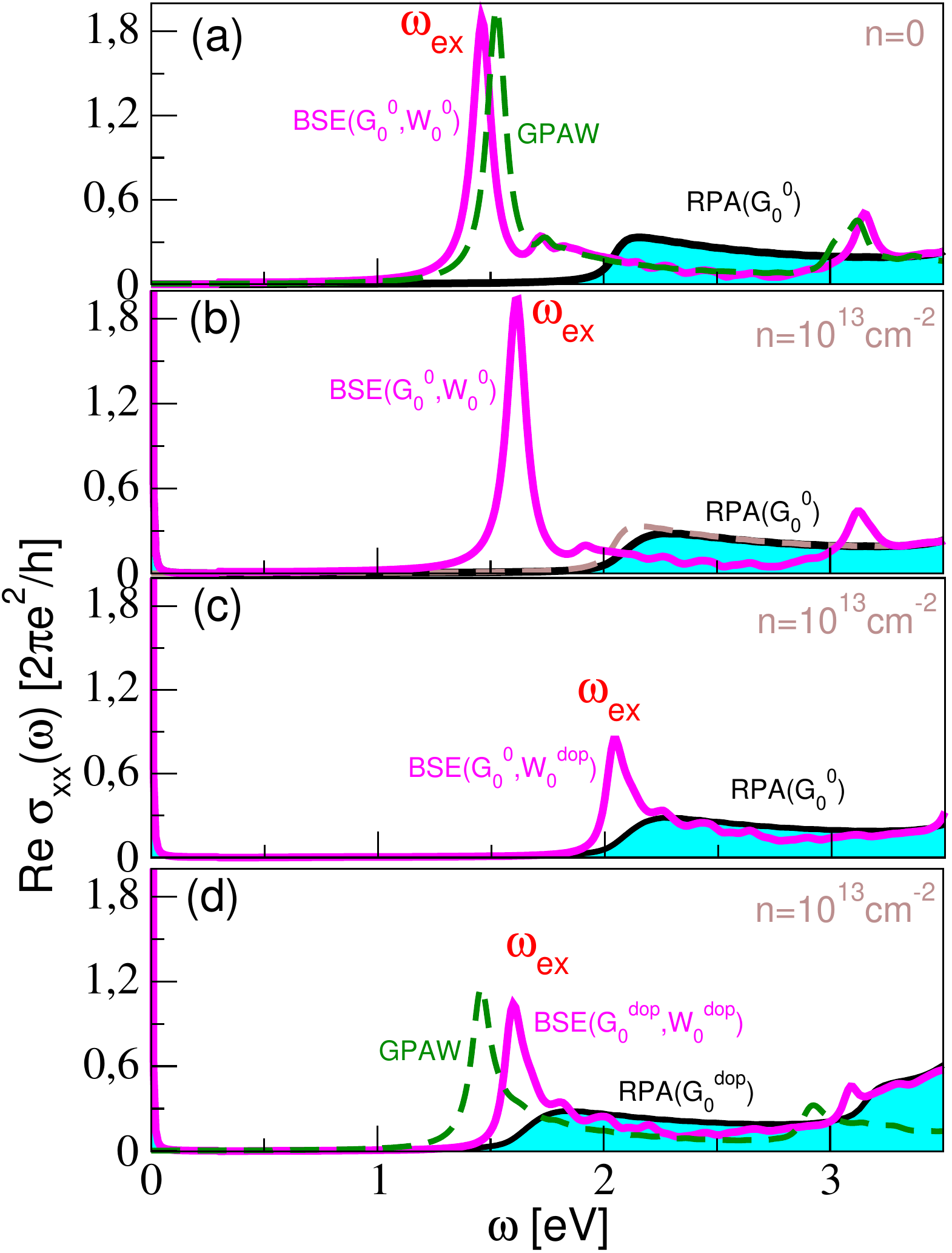}
\caption{The RPA (black) and  the `RPA+ladder' (magenta)  optical conductivities in (a) pristine and (b)-(d) doped phosphorene. (b) $G^0_0$  and $W^0_0$ are used at both the RPA and `RPA+ladder' level of calculations. (c) $G^0_0$ is used at RPA  and $G^0_0$ and $W^{dop}_0$ are used at the `RPA+ladder' level. (d) $G^{dop}_0$ and $W^{dop}_0$ are used, at both RPA and `RPA+ladder' level of calculations. 
The brown dashed line in panel (b) shows the RPA conductivity in pristine phosphorene for comparison.  In panels (b) and (c) the occupation factors $f_{n{\bf K}}$ appearing in the Green's function $G^0_0$ are taken to be as in a doped crystal. Green dashed lines in panel (a)  and (d) depict the result obtained by solving GW-BSE using the \textsc{gpaw} package.} 
\label{Fig6}
\end{figure}

\begin{figure}[b]
\centering
\includegraphics[width=7cm,height=5cm]{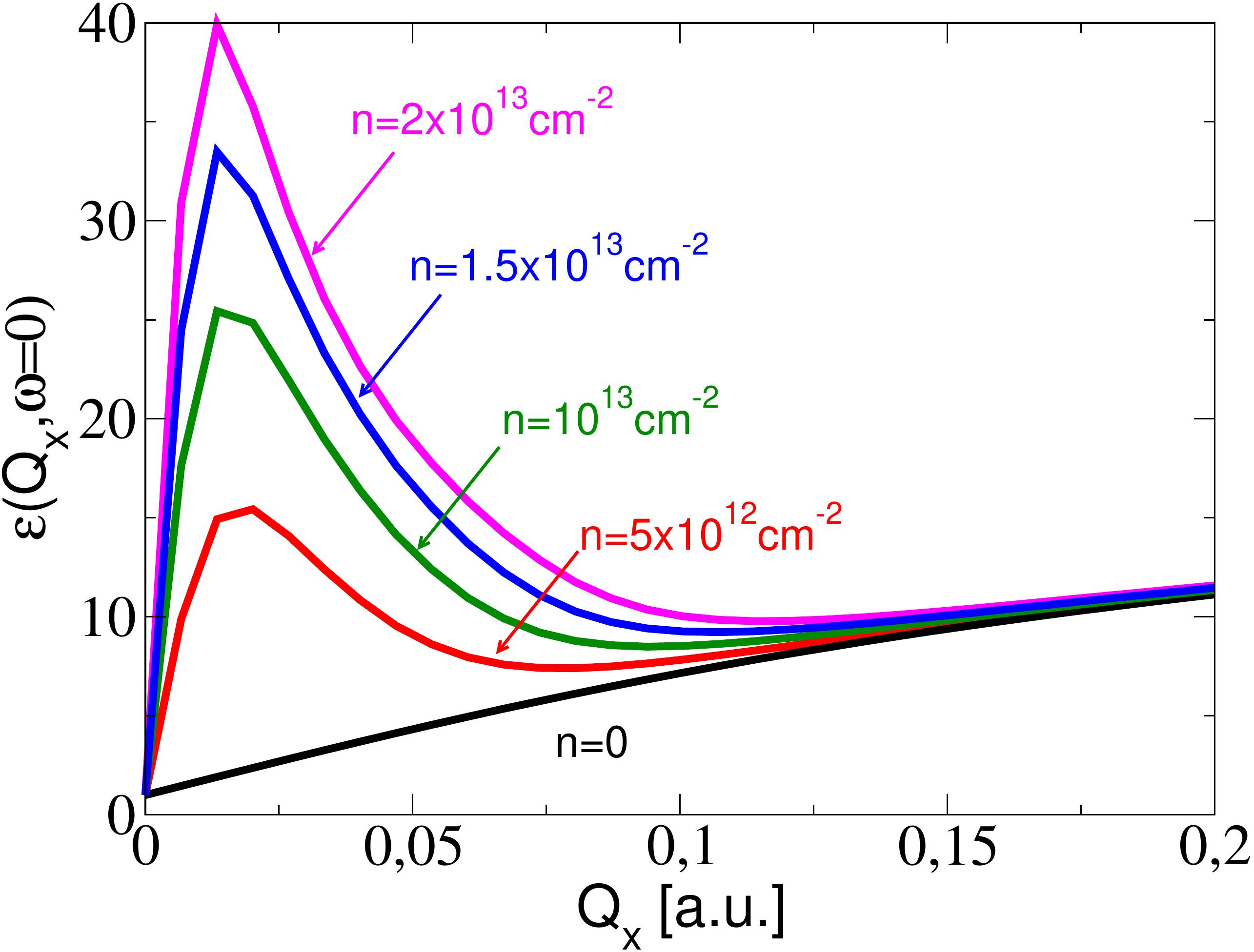}
\caption{The static dielectric function $\epsilon(Q_x,\omega=0)$ in doped phosphorene for different 
doping concentrations, i.e., $n=0$ (black), $n=5\times 10^{12}$cm$^{-2}$ (red),  $n=10^{13}$cm$^{-2}$ (green),  $n=1.5\times 10^{13}$cm$^{-2}$ (blue) and 
$n=2\times 10^{13}$cm$^{-2}$ (magenta).} 
\label{Fig7}
\end{figure}

Fig.\,\ref{Fig6}(b) shows the  RPA(G$^{0}_0$) and BSE($G^{0}_0,W^0_0$) optical conductivities in doped phosphorene, 
where $n=10^{13}$\,cm$^{-2}$. 
Here the pristine Green's function $G^0_0$ and the screened interaction $W^0_0$ are used at both the RPA and `RPA+ladder' level of 
calculation. However, since the goal is to explore what the impact of the Pauli blocking on the exciton spectral weight and the binding energy, the 
occupation factors $f_{n{\bf K}}$ which appear in the Green's functions $G^0_0$ are taken to be as in the doped sample. The same applies 
for the results presented in Fig.\,\ref{Fig6}(c). Thus, the only effect of the doping here is the extra population 
of the phosphorene conduction band $E_C$, which at $T=284$K shifts the Fermi energy by only $52$meV above $E_C$, as can be seen in Fig.\,\ref{Fig5}. 
The Pauli blocking reduces the phase space for direct interband electron-hole excitations and consequently blueshifts and reduces the intensity of the RPA absorption 
onset, which can be clearly seen when the black line is compared with the brown-dashed line showing the pristine RPA(G$^{0}_0$) conductivity. 
Consequently, the comparison between  BSE($G^{0}_0,W^0_0$) conductivities in Figs.\,\ref{Fig6}(a) and \ref{Fig6}(b) demonstrates how 
Pauli blocking affects the exciton energy. It can be noticed that the exciton is blue shifted to $\hbar\omega_{ex}=1.61$eV, such 
that its binding energy becomes $\Delta=544$\,meV. We can therefore conclude that the lack of phase space due to Pauli blocking 
reduces the exciton binding energy by $56$\,meV without affecting its oscillatory strength.  
Fig.\,\ref{Fig6}(c) shows the RPA(G$^{0}_0$) and BSE($G^{0}_0,W^{dop}_0$) optical conductivities. Here at the BSE stage of calculation, i.e. in the Fock kernel Eq.\ref{FockK}, the doped 
screened intaraction $W^{dop}_0$ is used. It can be noticed that an additional screening $\Delta W_0=W^{dop}_0-W^{0}_0$ significantly reduces the exciton binding 
energy and the oscilatory strength. More precisely, the exciton binding energy is reduced to $\Delta=114$\,meV. 
Interestingly, even such a small doping significantly changes the exciton 
identity, as even a small injection of charge carriers into the conduction band 
results in strong metallic screening that radically reduces the static interaction $W(Q,\omega=0)=v_Q/\epsilon(Q,\omega=0)$,
and thus the exciton binding energy and intensity. Fig.\,\ref{Fig7} shows the comparison between the  static dielectric function $\epsilon(Q_x,\omega=0)$ in 
pristine phosphorene (black)  and in the various cases of doped phosphorene (red, green, blue and magenta).  
While in the pristine phosphorene the dielectric function shows standard 
linear behavior $\epsilon(Q_x,\omega=0)=1+\alpha_x Q_x$, where $\alpha_x=68$, in the doped phosphorene it strongly overestimates the 
pristine value, especially in the long wave-length limit $Q\approx 0$. The same is valid for the $Q_y$ direction, where $\alpha_y=58$. 
Considering that $W(Q\approx 0)$ is exactly responsible for the formation of the electron-hole bound state it is not surprising that the exciton is significantly degraded.   
Finally, Fig.\,\ref{Fig6}(d) shows the RPA(G$^{dop}_0$) and the BSE($G^{dop}_0,W^{dop}_0$) optical conductivities where the total screened interaction 
$W^{dop}_0$, is used at both the RPA and `RPA+ladder' levels of calculations. 
Strong metallic screening $\Delta W^0$ reduces the band gap to $E_g= 1.58$\,eV, which also influences the exciton energy $\hbar\omega_{ex}=1.6$eV, as well as the 
exciton binding energy $\Delta=84$meV.
We emphasize here that the final drop in the exciton binding energy $\Delta$ of $516$\,meV and the
drop of the effective band gap $E_g+2\alpha(E_F-E_C)$ of $366$meV do not cancel, which 
results in a $150$\,meV blue shift of the exciton energy $\hbar\omega_{ex}$. 
Green  dashed line in Fig.\,\ref{Fig6}(d) shows the optical conductivity obtained by using the \textsc{gpaw} package. 
The qualitative agreement with RPA+ladder spectrum is still satisfactory good, however 
the \textsc{gpaw} exciton, at $\hbar\omega=1.47$eV  is $130$meV red shifted in comparison 
with  RPA+ladder exciton  providing its larger binding energy of $\Delta=214$meV. This disagreement is probably because the 
screened Colulomb interaction  (as shown in Fig.\ref{Fig7} very sensitive on small 
doping) is calculated using the RPA+ladder method more accuratelly (denser k-point mesh) than using  \textsc{gpaw} 
method. However, the \textsc{gpaw} result  still shows a small exciton blue shift of $20$\,meV in comparison with undoped case.     
The similar qualitative behaviour, exciton quenching and blue shift are also theoretically 
derived for case of doped single layer TMDs \cite{Exciton_vs_dop}.

\begin{figure}[t]
\centering
\includegraphics[width=8cm,height=11cm]{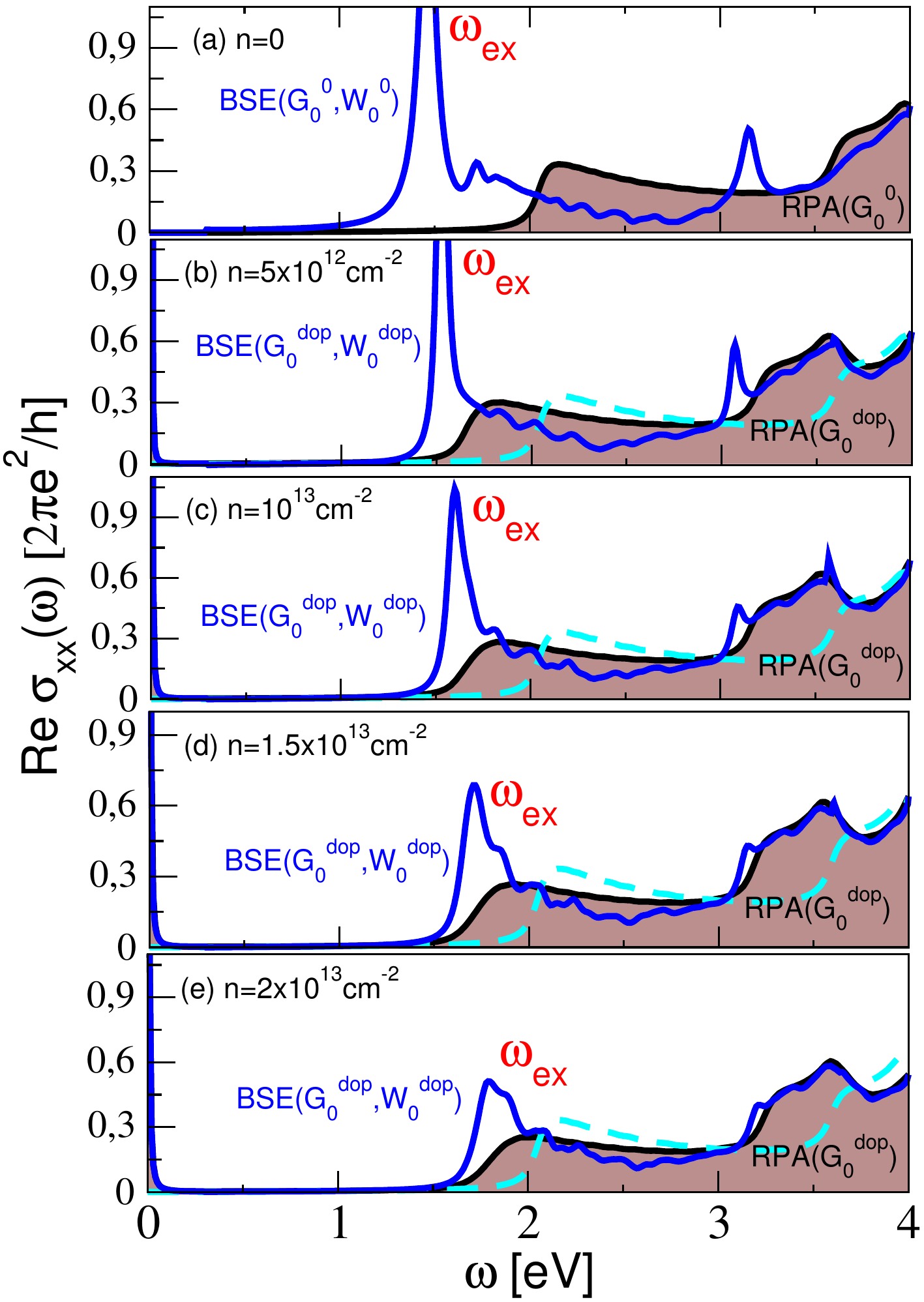}
\caption{Evolution of phosphorene exciton as a function of excess electron concentrations 
(a) $n=0$, (b) $n=5\times10^{12}$\,cm$^{-2}$, (c) $n=10^{13}$\,cm$^{-2}$, (d) $n=1.5\times10^{13}$\,cm$^{-2}$, and (e) $n=2\times10^{13}$\,cm$^{-2}$. 
The RPA(G$^{i}_0$) ($i=0,dop$) optical conductivities are shown with black and BSE($G^{i}_0,W^{i}_0$) ($i=0,dop$) optical conductivities 
by blue lines. For comparison, cyan dashed lines in panels (b)-(e) show the pristine RPA(G$^{0}_0$) optical conductivity.}
\label{Fig8}
\end{figure}

Figs.\,\ref{Fig8}(a)-(e) show the evolution of the phosphorene exciton as a function of excess electron concentration ($n>0$).
The RPA(G$^{i}_0$) ($i=0,dop$) optical conductivities are shown with the black lines and the BSE($G^{i}_0,W^{i}_0$) ($i=0,dop$) conductivities with the blue lines.  
For comparison, in Fig.\,\ref{Fig8}(a) we show the optical conductivities in pristine 
phosphorene ($n=0$), calculated from G$^{0}_0$ and W$^0_0$. 
In Figs.\,\ref{Fig8}(b)-(e) the optical conductivities for doped samples with
$n=5\times 10^{12}$\,cm$^{-2}$,  $n=10^{13}$\,cm$^{-2}$, $n=1.5\times 10^{13}$\,cm$^{-2}$ and $n=2\times 10^{13}$\,cm$^{-2}$, respectively, 
calculated from corresponding G$^{dop}_0$ and W$^{dop}_0$ are presented, while the cyan dashed line shows the  RPA(G$^{0}_0$) conductivity.
It can be clearly seen how excess electron concentration reduces the exciton binding energy and its 
oscillatory strength in addition to blueshifting the exciton energy $\hbar\omega_{ex}$. 
Quantitative values for the band gap $E_g$, exciton energy $\hbar\omega$, and exciton binding energy $\Delta$ (corresponding to Figs.\,\ref{Fig8}) are 
summarized in Table\,\ref{Table1}. It is clear that an already tiny electron doping of $n=5\times10^{12}$\,cm$^{-2}$ causes a drastic drop in the exciton 
binding energy, i.e., from $\Delta=600$\,meV to $128$\,meV. Further increase in the electron doping  
causes  weak additional decrease of the exciton binding energy. What is clearly noticeable 
from Table\,\ref{Table1} is that excess electrons cause a considerable blue shift of the exciton energy 
$\hbar\omega_{ex}$ such that, for example, already moderate electron doping of $n=2\times 10^{13}$cm$^{-2}$  
causes a blue shift of about $340$\,meV. This suggests that increasing doping causes a larger decrease in the exciton binding 
energy than does a decrease in the effective band gap $E_g+2(E_F-E_C)$.    
In Figs.\,\ref{Fig8}(b)-(e) the increasing intraband (or Drude) contribution to optical conductivity 
can be noticed in the THz ($\omega\approx 0$) region. The Drude contribution will be explained in more detail in Sec.\,\ref{Plpol}.

\begin{table}[!t] 
\begin{tabular}{c|c|c|c|c}
\hline
&&&
\\
$n$ [cm$^{-2}$] & \ \ $E_g$\ [eV]\ \ & \ \ $\hbar\omega_{ex}$ [eV]\ \ & $E_F-E_C$\ [meV]\ \  & $\Delta$[meV]
\\
&&&
\\
\hline
$0$& 2.05 & 1.45 & $/$  & $600$
\\
\hline
$0.5\times10^{13}$& 1.62  &1.53  &19  &  $128$
\\
\hline
$1.0\times10^{13}$&1.58 &1.6 &52 &  $84$
\\
\hline
$1.5\times10^{13}$&1.62 & 1.71&81 &   $72$
\\
\hline
$2.0\times10^{13}$&1.64 & 1.79&108 & $66$
\\
\hline
\end{tabular} 
\caption{Phosphorene band gap ($E_g$), exciton energy ($\hbar\omega_{ex}$), Fermi energy relative to conduction band ($E_F-E_C$), and exciton  binding energy ($\Delta$), 
according to Eq.\ref{ExBE}, for different doping concentrations $n$.}
\label{Table1}
\end{table}

\subsection{Exciton-polaritons}
In this section we explore the strength of hybridization between the phosphorene exciton and free-photons. 

\begin{figure*}[!t]
\includegraphics[width=0.4\textwidth]{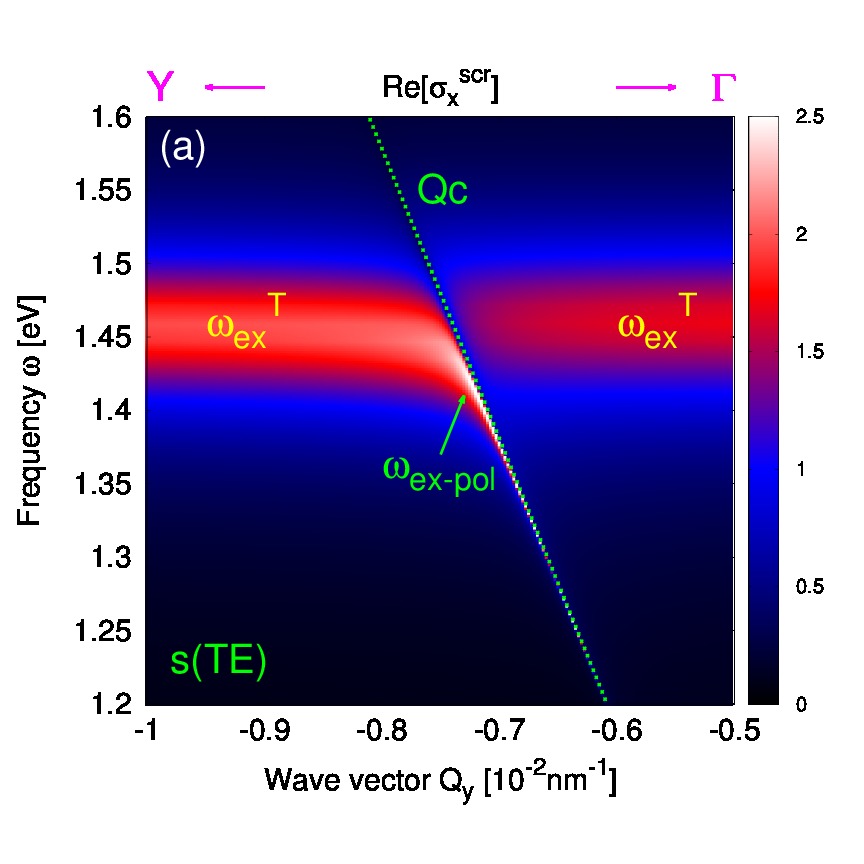}
\includegraphics[width=0.4\textwidth]{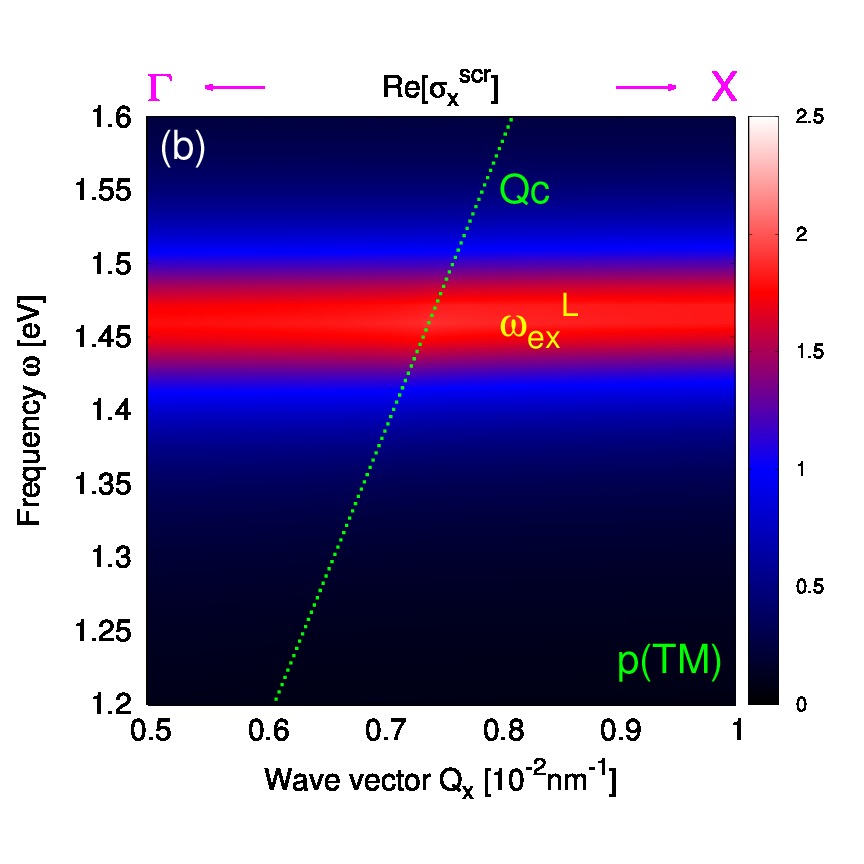}
\caption{Real part of the screened conductivity  (Re\,$\sigma^{scr}_{xx}$) in pristine phosphorene as a function of the transfer wave vector ${\bf Q}$ along (a) ${\bf Q}=Q_y{\bf y}$ and (b) ${\bf Q}=Q_x{\bf x}$ 
directions. Weak hybridization between the transverse exciton $\omega_{ex}^T$  and the photon $Qc$ forming exciton-polariton $\omega_{ex-pol}$ 
can be seen in panel (a).}
\label{Fig9}
\end{figure*}

Figs.\,\ref{Fig9}(a) and \ref{Fig9}(b) show the real part of the screened conductivity  Eq.\,\ref{ind-cur}  in pristine phosphorene as a function of the transfer wave vector ${\bf Q}$ along the ${\bf Q}=Q_y{\bf y}$ and ${\bf Q}=Q_x{\bf x}$ directions, respectively. The green dotted line represents the light-line $\omega=Qc$, i.e., the dispersion relation of 
free-photons. Therefore, Figs.\,\ref{Fig9}(a) and (b) actually show the intensities of transverse s(TE) and longitudinal p(TM) electromagnetic modes in pristine phosphorene, respectively. The intense pattern in Figs.\ref{Fig9}(a) in the evanescent region $\omega<Qc$ represents the intensity of the
evanescent transversal exciton $\omega^T_{ex}$ that hybridizes weakly with the free-photons as it approaches the light line $Qc$. It can be 
noticed that the  exciton intensity is enhanced and  slightly curved towards the light 
line $Qc$, indicating a certain hybridization with light and therefore the
formation of the exciton-polariton mode $\omega_{ex-pol}$. 
The intense signal which continues in the radiative region $\omega>Qc$ represents the radiative 
transverse exciton $\omega^T_{ex}$, the standard exciton seen in absorption spectra or in
photoluminescence  spectroscopy. It is of note that the radiative transverse exciton is of somewhat lower intensity than the evanescent 
transverse exciton. Fig.\,\ref{Fig9}(b) shows the intensity of the longitudinal exciton $\omega^L_{ex}$, which is dispersionless, and as expected does not interact with the transverse photons. Here we can conclude that the hybridization between 2D transverse excitons and free photons is quite 
weak and a stronger coupling may be achieved if the phosphorene is in the presence of a more confined electromagnetic field such as those produced by microcavity devices. 
A theoretical attempt to explain the exciton-polaritons in transition-metal dichalcogenides is given Ref.\,\cite{ex-pol3}
The hybridization between excitons in various TMDs and in microcavity electromagnetic modes has already been experimentally observed\,\cite{Nature_Polaritons,ex-pol1,ex-pol2,ex-pol4}.

\subsection{Plasmon-polaritons} 
\label{Plpol}
Here we present the intraband and interband RPA($G^{0,dop}_0$) conductivities, the effective number of charge 
carriers Eq.\,\ref{effectivencdc}, and the appearance of anisotropic plasmon-polaritons in pristine and doped phosphorene. 

\begin{figure*}[!t]
\includegraphics[width=0.4\textwidth]{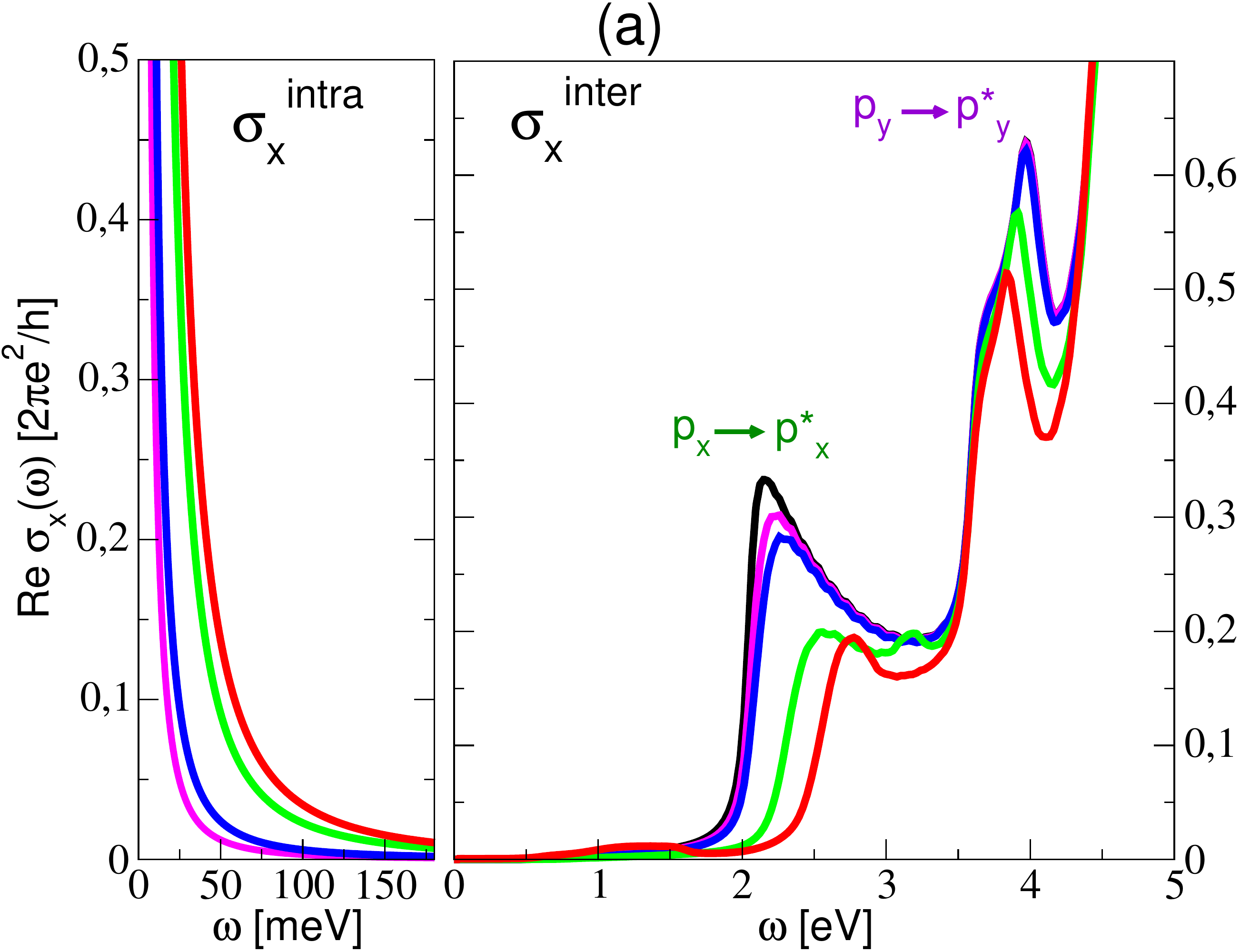}
\includegraphics[width=0.4\textwidth]{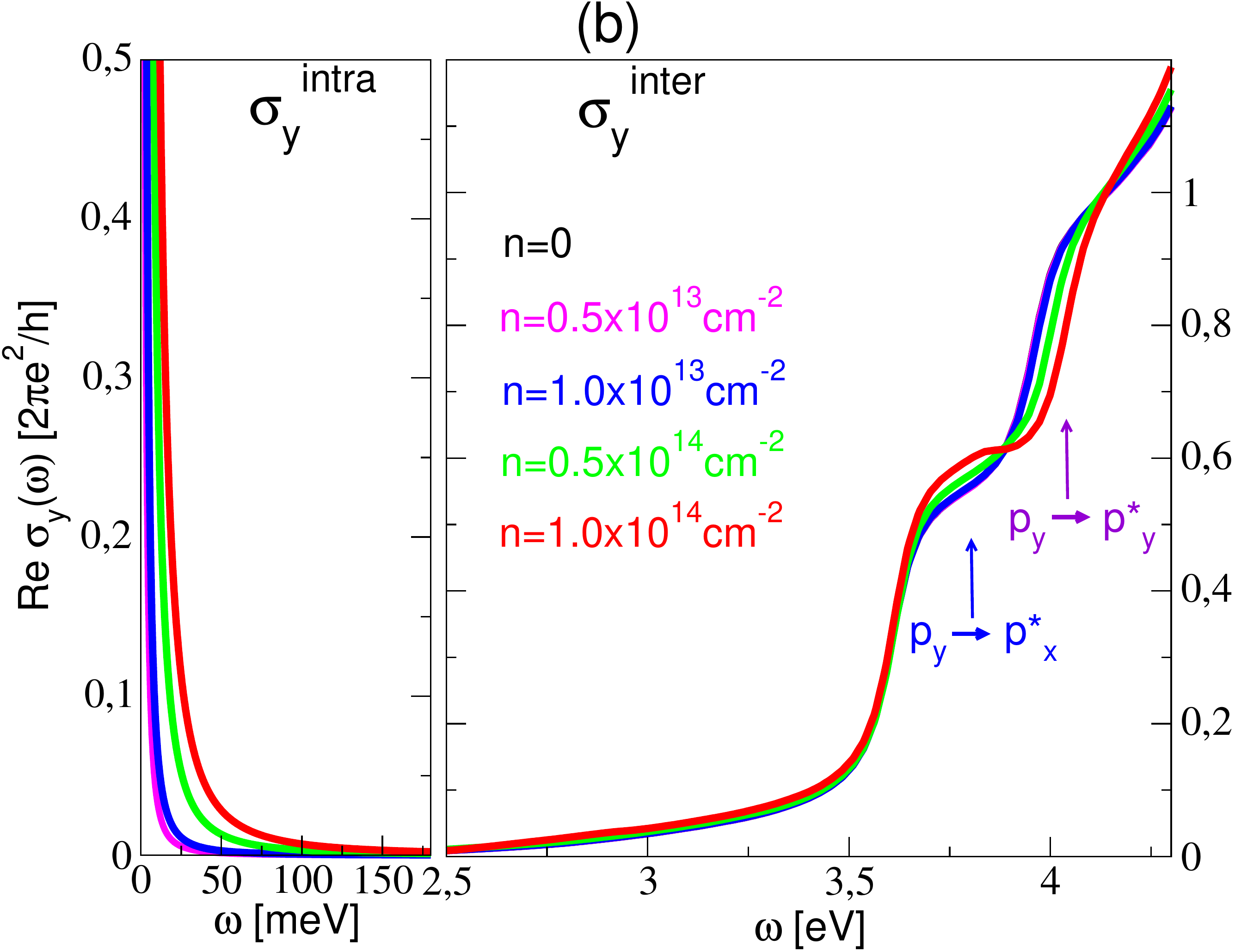}
\caption{RPA($G^{0,dop}_0$) optical conductivities (a) $\sigma_{xx}(\omega)$ and (b) $\sigma_{yy}(\omega)$ in 
doped phosphorene for various electron concentrations: $n=0$ (black),  $n=5\times10^{12}$\,cm$^{-2}$ (magenta),  $n=10^{13}$\,cm$^{-2}$ (blue),  
$n=5\times10^{13}$\,cm$^{-2}$ (green), $n=10^{14}$\,cm$^{-2}$ (red). Note the panels showing the separate intraband $\sigma_\mu^{intra}$ and interband $\sigma_\mu^{inter}$ contributions.}
\label{Fig10}
\end{figure*}

Figs.\,\ref{Fig10}(a) and \ref{Fig10}(b) show the RPA($G^{0,dop}_0$)  optical conductivities $\sigma_{xx}(\omega)$ and $\sigma_{yy}(\omega)$ in 
doped phosphorene for various electron concentrations: $n=0$ (black),  $n=5\times10^{12}$\,cm$^{-2}$ (magenta),  $n=10^{13}$\,cm$^{-2}$ (blue),  $n=5\times10^{13}$\,cm$^{-2}$ (green),  
$n=10^{14}$\,cm$^{-2}$ (red).
The interband contribution $\sigma^{inter}_x$ in pristine phosphorene ($n=0$) shows a characteristic onset which consists of a well defined asymmetric 
peak at $\omega\approx E_g$. This onset corresponds to $p_x\rightarrow p_{x}^*$ interband electron-hole excitations. 
At higher energies, namely at $\omega\approx 4$\,eV, another peak 
appears which corresponds to $p_y\rightarrow p_{y}^*$ interband electron-hole excitations, as seen in Fig.\,\ref{Fig5}.
When the electron concentration $n$ increases, the first peak $p_x\rightarrow p_{x}^*$ decreases and moves towards higher energies.  
As already discussed in Sec.\,\ref{OPTc}, this is a consequence of Pauli blocking, i.e., injected electrons occupy the bottom of the conductive band 
in the interval $E_F-E_C$ (as can be seen in Fig.\ref{Fig5}), which reduces the contribution of the direct interband electron-hole excitations 
in the energy interval $E_g<\omega<E_g+2(E_F-E_C)$, resulting in a blueshift of the peak of approximately $2(E_F-E_C)$.
The second peak $p_y\rightarrow p_{y}^*$ also decreases with doping, however, it is redshifted.
The interband contribution to conductivity $\sigma_y$ shows the lack of a strong peak at $\omega\approx E_g$.  
This is expected considering that the $y$ polarized light is not able to excite direct $p_x\rightarrow p_{x}^*$ excitations. 
The first step-like onset at $\omega\approx 3.7$\,eV corresponds to $p_y\rightarrow p_{x}^*$, and the second step-like onset at $\omega\approx 4$\,eV corresponds to the already mentioned 
$p_y\rightarrow p_{y}^*$ transitions. These onsets weakly depend on doping; the first onset slightly increases and redshifts, while 
second one decreases and blueshifts. The intraband/Drude conductivity  $\sigma^{intra}_\mu$ depends on the effective number of charge carriers $n^{e,h}_\mu$ (see Eq.\,\ref{effectivencdc})
which depends on the concentration of injected holes $n<0$ or electrons $n>0$ in the semiconductor. 
The effective number of charge carriers $n^{e,h}_\mu$, as shown later, finally defines the intensity 
of the plasmon-polariton. The left panels in Figs.\,\ref{Fig10}(a) and \ref{Fig10}(b) show how the increase of the excess electrons $n>0$ results in the increase of the Drude 
conductivities $\sigma_{x,y}$. Also, the Drude conductivity $\sigma_y$ is, for the same concentration $n$, smaller than the Drude conductivity $\sigma_x$. 

In order to analyze intraband conductivities $\sigma^{intra}_{x,y}$ quantitatively, in Table \ref{Table2} we list the effective 
concentrations of electrons and hole $n^{e,h}_x$ (second column) and $n^{e,h}_y$  (third column) for various doping concentrations $n$ (first column). 
The convention used here is: $n<0$ if the sample is doped by holes, and $n>0$ if the sample is 
doped by electrons. The $n^{e,h}_\mu$ ($\mu=x,y$)  are calculated using Eq.\,\ref{effectivencdc}, and the temperature is chosen to be $T=284$\,K. 
The fourth column shows the Fermi energy ($E_F$) in the doped sample relative to $E_V$ (if $n<0$) or $E_C$ (if $n>0$).
\begin{table}[!b] 
\begin{tabular}{c|c|c|c}
\hline
{\bf holes}&&&
\\
$n$[cm$^{-2}$] & \ \ $n^h_x\left[10^{-3}a_0^{-2}\right]$\ \ & \ \ $n^h_y\left[10^{-3}a_0^{-2}\right]$\ \  & $E_F-E_V$[meV]
\\
\hline
\ \ $-1\times10^{14}$& 14 & 1.9 &-339.6
\\
\hline
\ \ $-5\times10^{13}$&8.8 &0.83 &-165.8
\\
\hline
\ \ $-1\times10^{13}$& 2.2 &  0.125 &-18.5
\\
\hline
\ \ $-5\times10^{12}$&1.1  & 0.059 &7.1
\\
\hline
{\bf electrons}&&&
\\
$n$[cm$^{-2}$] & \ \ $n^e_x\left[10^{-3}a_0^{-2}\right]$\ \ & \ \ $n^e_y\left[10^{-3}a_0^{-2}\right]$\ \  & $E_F-E_C$[meV]
\\
\hline
$5\times10^{12}$&1.1 & 0.12&18.7
\\
\hline
$1\times10^{13}$&2.1 &0.25 &52.2
\\
\hline
$5\times10^{13}$& 8  &1.1   &254
\\
\hline
$1\times10^{14}$& 12 & 2.5 & $445$
\\
\hline
\end{tabular} 
\caption{Effective concentrations of holes $n^h_\mu$ and electrons $n^e_\mu$ as well as Fermi energies $E_F$ relative to the
valence $E_V$ or the conduction $E_C$ bands in doped phosphorene. The results are shown as a function of the doping concentration $n$, where $n<0$ corresponds to hole doping and $n>0$ corresponds to electron doping. The temperature is chosen to be $T=284$\,K.}
\label{Table2}
\end{table}
The decrease of excess holes ($n<0$) causes a decrease of the effective concentration of holes $n^h_x$, and an increase of injected electrons ($n>0$) causes an
increase of the effective concentration of electrons $n^e_x$, noting a symmetrical increase of concentrations $n^h_x$ and $n^e_x$ with respective increases in the 
concentrations $n<0$ and $n>0$, especially for small concentrations $|n|$. 
Somewhat different behavior applies to concentrations $n^h_y$ and  $n^e_y$.
These concentrations are (as also anticipated from Drude conductivities in  Figs.\,\ref{Fig10}) more than 10 times smaller than concentrations $n^{h,e}_x$. Also, the property of symmetrical increase is here violated, so that the concentration $n^e_y$ increases about twice as fast relative to the concentration 
$n^h_y$ (for smaller $|n|$). The effective concentrations $n^{e,h}_\mu$ define the intensity and frequency of collective modes arising due to hybridization between  
longitudinal 2D plasmons and photons, called plasmon-polaritons. 

\begin{figure*}[!t]
\includegraphics[width=0.33\textwidth]{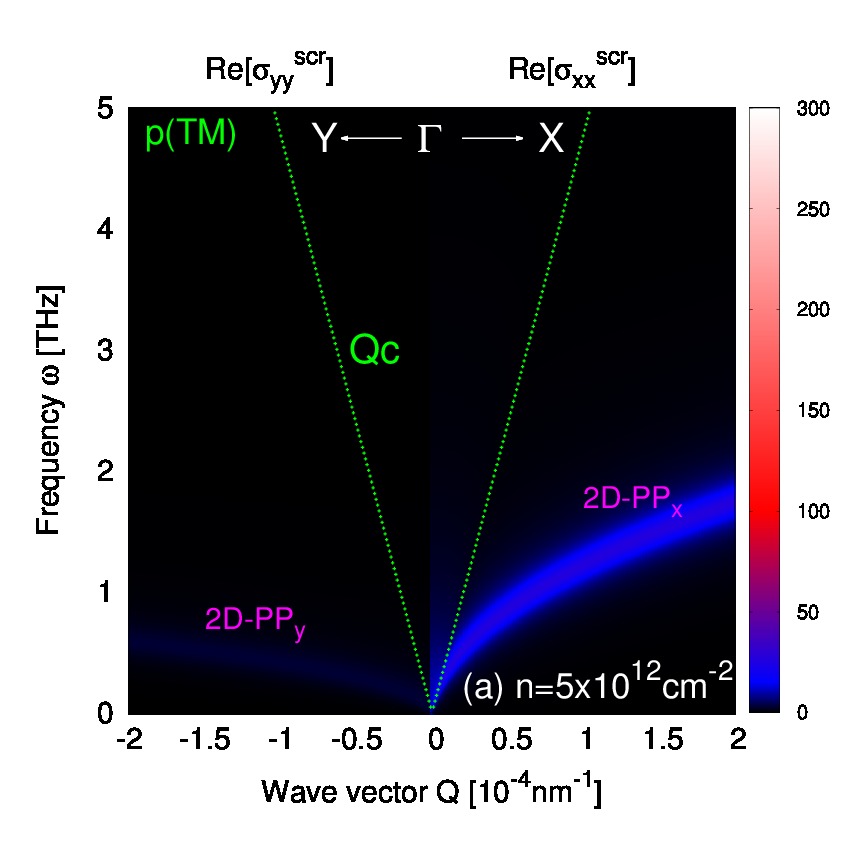}
\includegraphics[width=0.33\textwidth]{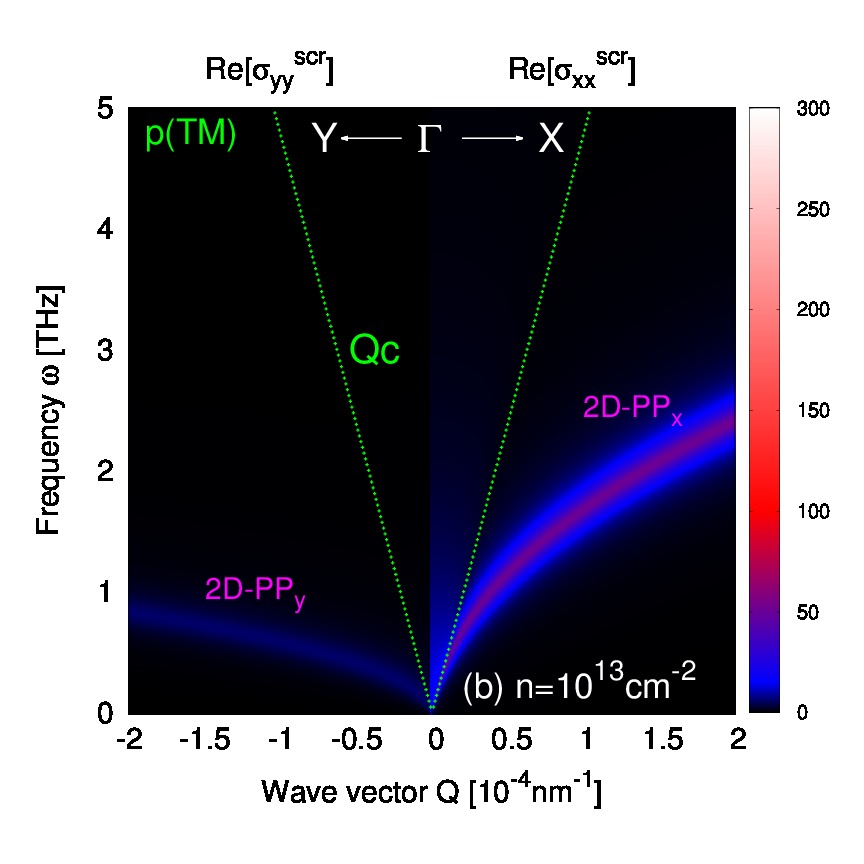}
\includegraphics[width=0.33\textwidth]{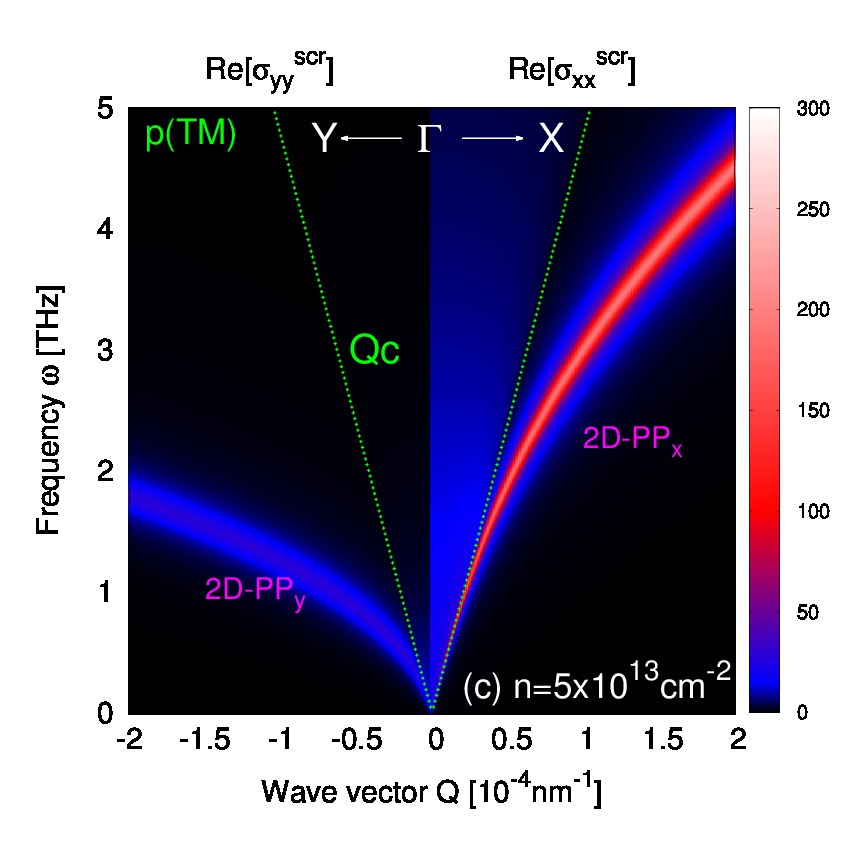}
\includegraphics[width=0.33\textwidth]{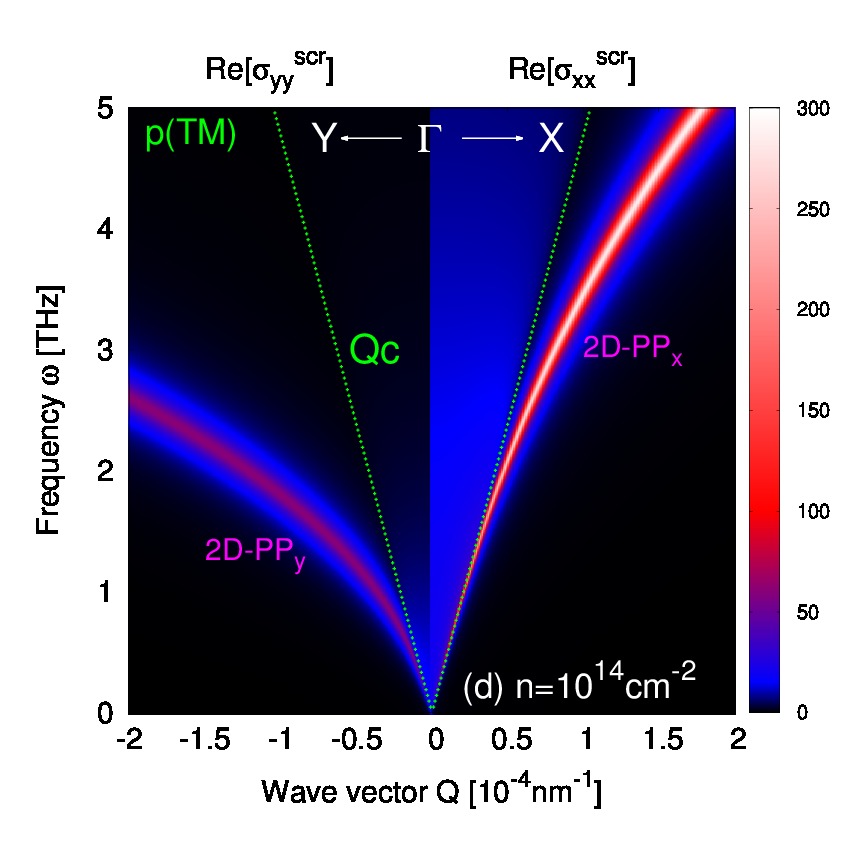}
\caption{Real part of the screened conductivities $\sigma^{scr}_{xx}$  and  $\sigma^{scr}_{yy}$ for momentum 
transfer ${\bf Q}=Q_x{\bf x}$ and  ${\bf Q}=Q_y{\bf y}$, respectively, as a function of doping concentrations: 
(a) $n=5\times 10^{12}$\,cm$^{-2}$, (b) $n=1\times 10^{13}$\,cm$^{-2}$, (c) $n=5\times 10^{13}$\,cm$^{-2}$ and  (d) $n=1\times 10^{14}$\,cm$^{-2}$. Panels show the 
intensities of longitudinal p(TM) electromagnetic modes in the THz  frequency region, i.e., the 2D plasmon polaritons 2D-PP$_x$ and 2D-PP$_y$.}
\label{Fig11}
\end{figure*}

Figs.\,\ref{Fig11}(a)-(d) show the real part of the screened conductivities Re\,$\sigma^{scr}_{xx}$ and Re\,$\sigma^{scr}_{yy}$ for 
momentum transfers $Q{\bf x}$ and  $Q{\bf y}$, respectively, in doped phosphorene as a function of excess electron concentrations, namely,  
(a) $n=5\times 10^{12}\,{\rm cm}^{-2}$, (b) $n=1\times 10^{13}\,{\rm cm}^{-2}$, (c) $n=5\times 10^{13}\,{\rm cm}^{-2}$ and (d) $n=1\times 10^{14}\,{\rm cm}^{-2}$.  
Momentum transfer $Q{\bf y}$ is here presented as a negative wave vector ($Q<0$). 
The polarization of induced currents is collinear with the direction of propagation and 
therefore Figs.\,\ref{Fig11}(a)-(d) represent the intensities of longitudinal p(TM) electromagnetic modes, i.e., 2D plasmon-polaritons 
2D-PP$_x$ and 2D-PP$_y$. We emphasize that the frequency scale here is in THz.   
Only the intraband conductivity $\sigma^{intra}_\mu$, or more precisely the effective numbers of charge carriers, determine the energy and the intensity 
of the 2D-PP$_\mu$. Therefore, following Eq.\,\ref{epsilonT}, the intense patterns seen in Figs.\,\ref{Fig11}(a)-(d) follow the zeros 
of the dielectric functions     
\begin{equation}
\epsilon_{\mu\mu}(Q\hat{\mu},\omega)=1+\frac{2\pi \beta L}{\omega}\sigma^{intra}_\mu(\omega);\ \ \mu=x,y.
\end{equation}
Consequently,  the $\sigma^{inter}$ and $\sigma^{ladd}$ do not affect plasmon-polaritons.  
As expected, by increasing the doping concentration $n>0$ and the concurrent increase in $n^e_\mu$, 2D-PP$_\mu$ become more intense 
and rise in energy. Also, the anisotropy in the effective concentrations $n^{e}_x>n^{e}_y$ is reflected in the anisotropy of 2D-PP$_\mu$ propagation, such that the 2D-PP$_x$ has larger energy and is more intense than 
2D-PP$_y$ for a given momentum value. It can be noticed that 2D-PP$_\mu$ only retains a polariton-like character (follows the light line $Qc$) 
at very small frequencies, soon after following the standard `square-root'-like 2D plasmon. 
However, polariton-like character increases gradually with doping 
$n$, so, for example, for dopings $n=1.0^{13}$\,cm$^{-2}$, $n=5.0^{13}$\,cm$^{-2}$ and $n=10^{14}$\,cm$^{-2}$ the 2D-PP$_x$ behaves as a polariton  up to $\omega<0.5$\,THz,   $\omega<1.0$\,THz, and $\omega<2.0$\,THz, respectively. Besides following the light line $Qc$ for very small $\omega$, 2D-PP$_x$ merges with the continuum of radiative electromagnetic modes, the blue pattern at $\omega>Qc$ that is
most noticeable in Fig.\,\ref{Fig11}(d). The merging with the continuum of radiative modes is considerably weaker for the $y$ polarised plasmon-polariton.
Here we can conclude that even a small fraction of the excess electrons in the phosphorene conduction band, ranging from $n\ \sim\ 5\times 10^{12}\ -\ 2\times 10^{13}$\,cm$^{-2}$ 
($E_F-E_C\ \sim\  19\ -\ 108$\,meV)  leads to a significant manipulation of the anisotropic plasmon-polariton intensity and energy.

\section{Conclusions}
\label{conclu}
We developed a formalism  suitable towards the study of electromagnetic modes in a wide class of conducting and semiconducting 2D
materials. The formulation can easily be adapted to calculate the electromagnetic modes in 2D van der Waals 
heterostructures or to calculate the interaction of these modes within confined cavity modes. Here the formulation was applied to calculate 
the optical conductivity (the evolution of the exciton intensity  and binding energy) in doped phosphorene. We have clearly demonstrated the mechanisms 
of exciton quenching (sudden drop of the exciton binding energy and intensity) due to injection of electrons in the phosphorene conduction band.    
Further, the formulation is applied to calculate the interaction of the phosphorene transverse exciton with free photons, where we have observed a weak hybridization and 
exciton-polariton formation. Finally, the method was applied to demonstrate the tuning of anisotropic plasmon-polaritons in phosphorene by electron doping.

\begin{acknowledgements}
V.D acknowledges financial support from Croatian Science Foundation (Grant no. IP-2020-02-5556) and European Regional Development Fund for the ``QuantiXLie Centre of Excellence'' (Grant KK.01.1.1.01.0004).
D.N. additionally acknowledges financial support from the Croatian Science Foundation (Grant no. UIP-2019-04-6869) and from the European Regional Development Fund for the ``Center of Excellence for Advanced Materials and Sensing Devices'' (Grant No. KK.01.1.1.01.0001). 
Computational resources were provided by the Donostia International Physic Center (DIPC) computing center as well as from the Imbabura cluster of Yachay Tech University, which was purchased under contract No. 2017-024 (SIE-UITEY-007-2017).
\end{acknowledgements}

\end{document}